\documentclass[lettersize,journal]{IEEEtran}
\IEEEoverridecommandlockouts
\usepackage{cite}
\usepackage[tbtags]{amsmath}
\usepackage{amssymb,amsfonts}
\usepackage{algorithm}
\usepackage{algorithmic}
\usepackage{graphicx}
\usepackage{textcomp}
\usepackage{xcolor}
\usepackage{caption}
\usepackage{subcaption}
\usepackage{commath}
\usepackage{url} 
\usepackage{multirow}

\newcommand{\overbar}[1]{\mkern 1.5mu\overline{\mkern-1.5mu#1\mkern-1.5mu}\mkern 1.5mu}

\def\BibTeX{{\rm B\kern-.05em{\sc i\kern-.025em b}\kern-.08em
    T\kern-.1667em\lower.7ex\hbox{E}\kern-.125emX}}

\newtheorem{theorem}{\textbf{Theorem}}

\begin{document}
\title{Low-Complexity ADMM-Based Multicast Beamforming in Cell-Free Massive MIMO Systems}

\author{Mahmoud Zaher, and Emil Björnson, \IEEEmembership{Fellow, IEEE,}%
\thanks{Mahmoud Zaher, and Emil Björnson are with the Department of Communication Systems, KTH Royal Institute of Technology, 164 40 Stockholm, Sweden (e-mail: mahmoudz@kth.se; emilbjo@kth.se).}

\thanks{This work was supported by the FFL18-0277 and SUCCESS grants from the Swedish Foundation for Strategic Research.}}

\maketitle

\begin{abstract}

The growing demand for efficient delivery of common content to multiple user equipments (UEs) has motivated significant research in physical-layer multicasting. By exploiting the beamforming capabilities of massive MIMO, multicasting provides a spectrum-efficient solution that avoids unnecessary intra-group interference. A key challenge, however, is solving the max-min fair (MMF) and quality-of-service (QoS) multicast beamforming optimization problems, which are NP-hard due to the non-convex structure and the requirement for rank-1 solutions. Traditional approaches based on semidefinite relaxation (SDR) followed by randomization exhibit poor scalability with system size, while state-of-the-art successive convex approximation (SCA) methods only guarantee convergence to stationary points. In this paper, we propose an alternating direction method of multipliers (ADMM)-based framework for MMF and QoS multicast beamforming in cell-free massive MIMO networks. The algorithm leverages SDR but incorporates a novel iterative elimination strategy within the ADMM updates to efficiently obtain near-global optimal rank-1 beamforming solutions with reduced computational complexity compared to standard SDP solvers and randomization methods. Numerical evaluations demonstrate that the proposed ADMM-based procedure not only achieves superior spectral efficiency but also scales favorably with the number of antennas and UEs compared to state-of-the-art SCA-based algorithms, making it a practical tool for next-generation multicast systems.

\end{abstract}


\section{Introduction}

The growing demand for wireless services and multimedia applications continues to push the capacity limits of communication networks. Physical-layer multicasting provides an efficient transmission technique by mitigating intra-group interference and improving spectrum utilization. Through beamforming, multicasting can simultaneously deliver the same information to multiple user equipments (UEs) within a coverage area using a single transmission.
It enables diverse applications in modern wireless networks. For example, mobile operators can support simultaneous video streaming of live events or videoconferencing to many UEs as an efficient alternative to unicast-based content delivery \cite{hsu2017joint}. It is also well-suited for services such as the distribution of machine learning models in federated learning, where multiple UEs require identical data. In addition, multicasting is a natural choice for emergency alert systems, enabling critical information to be disseminated quickly and reliably to large groups of UEs. The primary challenge with multicasting is then to optimize the common multicast precoder in order to boost the desired signal of all UEs simultaneously, which is the main focus of this paper.

Massive multiple-input multiple-output (MIMO) has emerged as a key enabler for efficient multicast beamforming. However, conventional cell-centric architectures remain limited by inter-cell interference and uneven pathloss \cite{zaher2023soft}. Cell-free massive MIMO, where geographically distributed access points (APs) cooperate without cell boundaries, has attracted significant interest due to its ability to provide uniform coverage and enhance the performance of cell-edge UEs \cite{bjornson2019making,demir2021foundations,zaher2023learning}. This property is particularly relevant to multicast transmissions, where group performance is dictated by the weakest UE \cite{karipidis2008quality}. Moreover, a serious impediment to the deployment of cell-free massive MIMO systems lies in fronthaul capacity limitations. The main fronthaul load comes from distributing the data from the central processing unit (CPU) to the APs. Another advantage of multicasting is that it reduces the fronthaul requirements compared to unicasting since all UEs in a multicast group are served using a single transmission.

Multicast beamforming optimization has been studied under quality-of-service (QoS) and max–min fairness (MMF) criteria, where the QoS level ensures a minimum received SNR for all UEs, and MMF aims to maximize the lowest received SNR. The foundational works \cite{sidiropoulos2006transmit,karipidis2008quality} established that the problem is NP-hard and proposed semidefinite relaxation (SDR) with randomization to extract feasible rank-1 solutions. However, this approach scales poorly as it suffers from degraded approximation quality when the number of transmit antennas and UEs grows large \cite{ashraphijuo2017multicast,zhou2017coordinated}.  Moreover, in the case of having more than one multicast group, a multi-group multicast power control (MMPC) optimization problem needs to be solved for each candidate beamformer obtained through randomization, requiring extensive computations. Note that a rank-1 beamforming solution, where a single multicast signal is transmitted for each multicast group using spatial multiplexing, represents a practical choice due to its ease of implementation compared to a higher-rank transmission.

In the context of massive MIMO multicasting, asymptotic beamforming structures have been investigated. For example, \cite{xiang2014massive} derived optimal non-cooperative beamformers as linear combinations of channel vectors, while \cite{sadeghi2017reducing} proposed a two-layer design combining low-complexity inter-group interference suppression and intra-group successive convex approximation (SCA)-based refinement. Conventional maximum ratio (MR) and regularized zero-forcing (RZF) precoding for the composite multicast channel with a pilot power allocation scheme that compensates for the pathloss differences between UEs in a multicast group and the serving BS have been explored in \cite{sadeghi2017max,de2022user,li2022spectral}. These designs exploit asymptotic channel properties but converge slowly for the case of multicast beamforming as the number of UEs per group grows, highlighting the performance gap between unicast and multicast for such asymptotic designs.
In \cite{dong2020multi,zhang2023ultra}, the optimal multicast beamforming structure is exploited to provide a reduction in computational complexity when the number of transmit antennas is much greater than the number of UEs, by relying on approximations to the involved parameter matrices of the optimal structure.

To overcome these limitations, several iterative optimization techniques have been proposed. Low-complexity solutions based on weak Lagrangian duality \cite{dartmann2011low} and iterative second-order cone programming (SOCP) \cite{tran2013conic}  have addressed single-multicast group and single-cell scenarios with a limited number of UEs per cell. In \cite{zhou2017coordinated}, a multi-group multicast MMF problem is formulated and solved using parametric manifold optimization in a multicell network with a single multicast group per cell. The difference-of-convex approximation (DCA) algorithm in \cite{hsu2017joint} extend to multicell setups and leverages SCA, which has become the state-of-the-art in multicast beamforming optimization. In our primal work \cite{globecom_paper,zaher2026cell}, we proposed a new optimization procedure, namely the successive elimination algorithm (SEA), that relies on SDR followed by iterative elimination of higher-rank solutions to extract a near-optimal rank-1 solution to the MMF multicast problem, showing superior performance in terms of SE and computational complexity to the SDR with randomization and SCA-based methods when standard solvers were utilized. Other works considered weighted sum rate maximization \cite{wang2016weighted}, although QoS and MMF remain more suitable objectives since the group rate is bottlenecked by the weakest UE \cite{park2008capacity}.

Despite the performance gains provided by SCA- and SEA-based methods, the high computational complexity for multicast beamforming optimization remains a serious impediment to the practicality of such beamforming designs, even for medium-sized problems. To reduce the complexity of multicast beamforming optimization, a recent line of work \cite{zhang2023ultra,chen2017admm,konar2017fast,ibrahim2020fast,mohamadi2022low,mohamadi2024low} has developed different first-order methods (FOMs) to replace general-purpose solvers to achieve a low-complexity local optimum solution to the QoS and/or MMF multicast problems in different cellular network setups. In general, each algorithm is tailored for a specific optimization objective and SCA formulation, requiring a new algorithm for each problem. In \cite{huang2016consensus}, the proposed consensus alternating direction method of multipliers (ADMM) algorithm requires many auxiliary variables, which increases the computational complexity and does not provide convergence guarantees due to the non-convexity of the multicast beamforming problem. In \cite{konar2017fast}, several SCA-based FOMs are developed, where an inexact version of ADMM that utilizes proximal operators to update the primal variables, namely Linearized-ADMM, shows consistently superior performance. The prior state-of-the-art SCA-ADMM algorithm focusing solely on multicast beamforming optimization is proposed in \cite{chen2017admm}. A similar SCA-ADMM algorithm has been utilized in \cite{zhang2023ultra}. The algorithm shows superior performance compared to algorithms that rely on the consensus ADMM approach. The authors in \cite{ibrahim2020fast} utilize similar FOMs to solve the problem of joint multicast beamforming optimization and antenna selection. The work in \cite{mohamadi2022low} tackles the problem of robust multicast beamforming design with a two-layer ADMM algorithm, in which the inner layer relies on consensus ADMM, whereas \cite{mohamadi2024low} extends the solution to incorporate antenna selection.
The conference version of this paper \cite{ADMM_conference_paper} devised the first ADMM algorithm for our state-of-the-art SDR-based SEA to solve the MMF and QoS problems limited to a single-cell single-power constraint setup, demonstrating its strong potential for improving efficiency in this field.

In this paper, we tackle the multicast beamforming optimization problem in cell-free massive MIMO with per-AP power constraints, which represents a generalization to the conference version. We derive a new formulation for the multicast MMF and QoS optimization problems and utilize a novel approach to devise a fast ADMM-based solution in this extended scenario. Different from our prior work \cite{globecom_paper,zaher2026cell}, which relies on standard solvers, the proposed SEA-ADMM iterative optimization procedure achieves near-global optimal beamforming solutions to the MMF and QoS objectives at a vastly reduced computational complexity. A key advantage of our approach is its adaptability to various multicast optimization objectives, network architectures and configurations. Throughout the paper, we focus on multicast beamforming design, assuming perfect CSI at the cell-free network side. The main contributions are summarized as follows:
\begin{itemize}
\item We introduce new MMF and QoS problem formulations that are key to developing a fast and general SDP-ADMM optimization framework for multicast beamforming. Moreover, the new formulations directly consider per-AP power budgets, eliminating the redundant total power constraint that is commonly found in previous literature, thereby reducing complexity.

\item We propose a novel ADMM-based iterative elimination strategy coupled with SDR to efficiently obtain near-global optimal rank-1 beamforming solutions to the MMF and QoS multicast problems with per-AP power constraints. The proposed algorithm with successive rank reduction represents the first SDP-ADMM algorithm in the context of multicell multicast beamforming optimization and is capable of yielding near-global optimal solutions, surpassing prior state-of-the-art SCA methods that can only converge to stationary points of the non-convex problems.
We highlight that the proposed ADMM elimination procedure can be employed as an effective low-complexity rank reduction method not only for multicast beamforming but also for a wide class of optimization problems that utilize SDR or SCA methods.

\item We provide extensive numerical evaluations demonstrating significant gains in SE and computational complexity over existing SDR- and SCA-based multicast beamforming optimization methods.
\end{itemize}

The rest of the paper is organized as follows: Section~\ref{sys_mod} presents the multicast communication system model and optimization objectives. Section~\ref{successive_elimiation_sec} presents the core concepts of the SEA. In Section~\ref{SEA_ADMM}, the proposed ADMM iterative optimization procedure for the MMF and QoS problems is detailed. Section~\ref{results} presents numerical results, whereas the main conclusions are summarized in Section~\ref{conc}.

\textbf{Notations:} Lowercase and uppercase boldface letters denote column vectors and matrices, respectively. The symbols $(\cdot)^*$, $(\cdot)^T$, and $(\cdot)^H$ indicate conjugate, transpose, and conjugate transpose, respectively. $\mathbb{E}(\cdot)$, tr$(\cdot)$, $\norm{\cdot}_p$, and $\norm{\cdot}_F$ denote the expectation, trace, $l_p$ vector norm and Frobenius matrix norm, respectively. $\mathbf{I}_M$ represents the $M \times M$ identity matrix.

\section{System Model} \label{sys_mod}

\begin{figure}
\centering
\setlength{\abovecaptionskip}{0.33cm plus 0pt minus 0pt}
\includegraphics[scale=0.65]{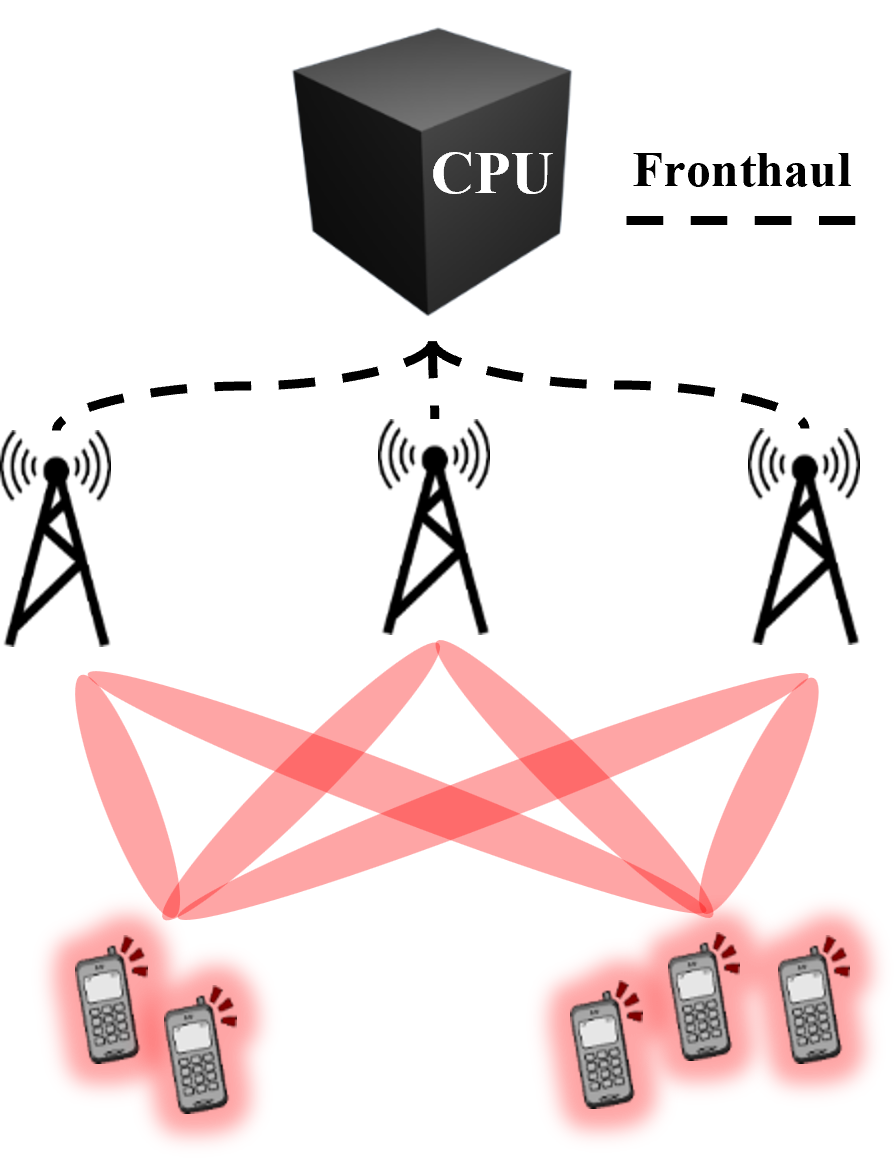}
\caption{Irregular cell-free massive MIMO multicast beamforming to a UE group.}
\label{system_model_fig}
\end{figure}

In this paper, we consider a cell-free massive MIMO network with $L$ APs, each equipped with $N$ antennas. The APs jointly serve $K$ single-antenna UEs that are arbitrarily distributed in a large service area, with a single multicast transmission. We consider a narrow-band channel, such that each channel realization is frequency-flat and quasi-static in time.\footnote{The proposed solution can be applied to each subcarrier in a multi-carrier system.} The channel realizations are assumed to be available at the APs.
The channel between UE $k$ and AP $l$ is denoted as $\mathbf{h}_{kl} \in \mathbb{C}^N$. The system model is depicted in Fig. \ref{system_model_fig}.

The received signal at UE $k$ is computed as
\begin{equation}
    y_{k}^\mathrm{dl} = \sum_{l = 1}^L\mathbf{h}_{kl}^H\mathbf{w}_ls + n_{k},
\end{equation}
where $s$ denotes the zero-mean unit-variance multicast signal intended for all UEs, $\mathbf{w}_l$ represents the common multicast precoding vector of AP $l$, and $n_{k} \sim \mathcal{N}_{\mathbb{C}}(0, \sigma_{k}^2)$ is the noise at UE $k$. The multicast precoding vector for each AP satisfies a short-term power constraint, which means that the power constraint must be satisfied for each channel realization. Accordingly, $\norm{\mathbf{w}_l}_2^2 \leq P_{l,\mathrm{max}}$, where $P_{l,\mathrm{max}}$ represents the maximum transmit power of AP $l$. As a result, the achievable SE of UE $k$ under the perfect CSI assumption is
\begin{equation}
    \textrm{SE}_{k}^{\mathrm{dl}} = \textrm{log}_2\left(1 + \textrm{SNR}_k^{\mathrm{dl}}\right).
    \label{SE}
\end{equation}

For notational convenience, we define the concatenated channel vector between all APs and UE $k$ as $\mathbf{h}_{k} = \left[\mathbf{h}_{k1}^T, \mathbf{h}_{k2}^T, \hdots,  \mathbf{h}_{kL}^T\right]^T \in \mathbb{C}^{LN \times 1}$ and the corresponding precoding vector as $\mathbf{w} = \left[\mathbf{w}_{1}^T, \mathbf{w}_{2}^T, \hdots, \mathbf{w}_{L}^T\right]^T \in \mathbb{C}^{LN \times 1}$.
Further, we define the notation $\mathbf{H}_k = \mathbf{h}_k\mathbf{h}_k^H/\sigma_k^2$ and $\mathbf{W} = \mathbf{w}\mathbf{w}^H$. Utilizing the fact that $\left|\mathbf{h}_{k}^H\mathbf{w}\right|^2 = \textrm{tr}\left(\mathbf{h}_{k}\mathbf{h}_{k}^H\mathbf{w}\mathbf{w}^H\right)$, the SNR of UE $k$ in \eqref{SE} can be written as
\begin{equation}
    \textrm{SNR}_k^{\mathrm{dl}} = \frac{\left|\mathbf{h}_{k}^H\mathbf{w}\right|^2}{\sigma_{k}^2} = \textrm{tr}\left(\mathbf{H}_k\mathbf{W}\right).
    \label{SNR}
\end{equation}

In this paper, we focus on the MMF and QoS multicast problems. An advantage of multicasting is that there is no inter-user interference, since all UEs are served using a single transmission. For the MMF objective, the goal is to find the multicast transmit precoding vector $\mathbf{w}$ that maximizes the minimum achievable SNR among all UEs in the system under the per-AP power constraints. Conversely, the QoS objective aims to minimize the per-AP powers subject to a given SNR target. In a multicast transmission, the minimum SNR among the UEs determines what data rate can be utilized since all UEs must be able to decode the same signal. Hence, the MMF and QoS objectives represent appropriate design criteria for the multicast problem. The SNR formulation in \eqref{SNR} is suitable to construct the SDP of the MMF and QoS multicast problems.

\subsection{Max-Min Fair Multicast Problem} \label{max_min_sec}

We define $\mathbf{D}_l = \textrm{blkdiag}\left(\mathbf{Z}_{l1}, \hdots, \mathbf{Z}_{lL}\right)$, where $\textrm{blkdiag}(\cdot)$ denotes a block-diagonal matrix with $\mathbf{Z}_{ll} = \mathbf{I}_N$ and $\mathbf{Z}_{li} = \mathbf{0}_{N \times N}, \forall i \neq l$. Hence, the MMF multicast problem with per-AP power constraints can be written as
\begin{subequations}
\begin{align}
    \mathop{\mathrm{maximize}}\limits_{\mathbf{W}} &\hspace{-3pt}\min_{\hspace{0.5em}\substack{ k \in \{1, \hdots, K\} }}\hspace{0.2em} \textrm{tr}\left(\mathbf{H}_{k}\mathbf{W}\right) \\
    \textrm{s.t.} &\quad \textrm{tr}\left(\mathbf{D}_l\mathbf{W}\right) \leq P_{l,\mathrm{max}}, \quad \forall l \in \{1, \hdots, L\},\\
    &\quad \mathbf{W} \succeq \mathbf{0},
    \quad \textrm{rank}\left(\mathbf{W}\right) = 1.\label{rank1_MMF}
\end{align}
\label{MMF_problem}
\end{subequations}

Due to the non-convex rank-1 constraint in \eqref{rank1_MMF}, the above problem is non-convex. By dropping the non-convex rank-1 constraint, the relaxed convex counterpart of the problem is devised. Moreover, the problem is cast in epigraph form by introducing an auxiliary variable $t \geq 0$ to lower bound the worst-case SNR. Accordingly, the relaxed MMF problem is reformulated as
\begin{subequations}
\begin{align}
    \mathop{\mathrm{maximize}}\limits_{\mathbf{W},\textrm{ }t} &\quad t \\
    \textrm{s.t.} &\quad \textrm{tr}\left(\mathbf{H}_{k}\mathbf{W}\right) \geq t, \quad \forall k \in \{1, \hdots, K\}, \\
    &\quad \textrm{tr}\left(\mathbf{D}_l\mathbf{W}\right) \leq P_{l,\mathrm{max}}, \quad \forall l \in \{1, \hdots, L\},\\
    &\quad \mathbf{W} \succeq \mathbf{0}.
\end{align}
\label{MMF_problem2}
\end{subequations}

Let $\mathbb{H}^N$ denote the set of $N \times N$ Hermitian matrices. The linear maps $\mathcal{H}\left(\cdot\right): \mathbb{H}^N \rightarrow \mathbb{R}^K$ and $\mathcal{D}\left(\cdot\right): \mathbb{H}^N \rightarrow \mathbb{R}^L$ are defined as
\begin{align}
    &\mathcal{H}\left(\mathbf{W}\right) = \left(\langle\mathbf{H}_1^H, \mathbf{W}\rangle, \hdots, \langle\mathbf{H}_K^H, \mathbf{W}\rangle\right),\\
    &\mathcal{D}\left(\mathbf{W}\right) = \left(\langle\mathbf{D}_1^H, \mathbf{W}\rangle, \hdots, \langle\mathbf{D}_L^H, \mathbf{W}\rangle\right),
\end{align}
where the inner product between two matrices is given by $\langle\mathbf{A}, \mathbf{B}\rangle = \textrm{tr}\left(\mathbf{A}^H\mathbf{B}\right)$. The relaxed MMF problem can then be written compactly as
\begin{subequations}
\begin{align}
    \mathop{\mathrm{maximize}}\limits_{\mathbf{W},\textrm{ }t} &\quad t \\
    \textrm{s.t.} &\quad \mathcal{H}\left(\mathbf{W}\right) \geq \mathbf{1}t, \label{MMF_SNR} \\
    &\quad \mathcal{D}\left(\mathbf{W}\right) \leq \mathbf{p}, \label{MMF_pow}\\
    &\quad \mathbf{W} \succeq \mathbf{0}, \label{MMF_psd}
\end{align}
\label{MMF_problem3}%
\end{subequations}
where $\mathbf{p} = [P_{1,\mathrm{max}}, \hdots, P_{L,\mathrm{max}}]^T$ and $\mathbf{1} \in \mathbb{R}^K$ is the all one vector.

\subsection{Quality-of-Service Multicast Problem}

In contrast to the MMF objective, the QoS objective aims to guarantee a given SNR target for all multicast UEs while minimizing the per-AP powers. Hence, the QoS multicast problem can be written as
\begin{subequations}
\begin{align}
    \mathop{\mathrm{minimize}}\limits_{\mathbf{W}} &\hspace{-3pt}\max_{\hspace{0.5em}\substack{ l \in \{1, \hdots, L\} }}\hspace{0.2em} \frac{1}{P_{l,\mathrm{max}}}\textrm{tr}\left(\mathbf{D}_l\mathbf{W}\right) \\
    \textrm{s.t.} &\quad \textrm{tr}\left(\mathbf{H}_{k}\mathbf{W}\right) \geq \gamma_k, \quad \forall k \in \{1, \hdots, K\},\\
    &\quad \mathbf{W} \succeq \mathbf{0},
    \quad \textrm{rank}\left(\mathbf{W}\right) = 1.\label{rank1_QoS}
\end{align}
\label{QoS_problem}%
\end{subequations}
where $\boldsymbol{\gamma} = [\gamma_1, \hdots, \gamma_K] \in \mathbb{R}^{K}$ represents the SNR targets of the UEs. As in the case of the MMF problem, the relaxed convex QoS problem is devised by dropping the non-convex rank-1 constraint. Further, the problem is cast in epigraph form by introducing an auxiliary variable $x$ to upper bound the per-AP powers. The relaxed QoS problem can then be reformulated as
\begin{subequations}
\begin{align}
    \mathop{\mathrm{minimize}}\limits_{\mathbf{W},\textrm{ }x} &\quad x \\
    \textrm{s.t.} &\quad \mathcal{H}\left(\mathbf{W}\right) \geq \boldsymbol{\gamma}, \\
    &\quad \mathcal{D}\left(\mathbf{W}\right) \leq \mathbf{p}x,\\
    &\quad \mathbf{W} \succeq \mathbf{0}.
\end{align}
\label{QoS_problem2}
\end{subequations}

In general, such relaxation results in the solution matrix $\mathbf{W}$ having a high rank that does not satisfy the rank-1 constraint of the original MMF and QoS problems, requiring post-processing of the output matrix to extract a feasible rank-1 beamforming solution. Finding the optimal post-processing is challenging.

\section{Successive Elimination Algorithm}\label{successive_elimiation_sec}

Standard SDP solvers tend to prefer higher-rank solutions over lower-rank ones, even when both achieve the same objective value \cite{luo2010semidefinite}. The reason is that low-rank solutions to SDP problems contain many zero eigenvalues and so typically lie on the boundary of the feasible set rather than in its interior, since a small perturbation to these eigenvalues may push the solution to be infeasible. As a result, the relaxed SDP problem often yields high-rank solutions, even when a rank-1 solution exists. Prior works using SDR have commonly applied randomization techniques \cite{karipidis2008quality} to obtain rank-1 beamforming solutions from the higher-rank solutions. However, these approaches scale poorly; as the total number of antennas and UEs increase, a significantly larger number of random candidate beamforming vectors is required to maintain good SE performance. 
To address this limitation, we adopt the state-of-the-art SEA introduced in our earlier work \cite{globecom_paper,zaher2026cell}.

\subsection{Penalty Design}

The SEA iteratively eliminates higher-rank solutions to the SDP that yield an optimum objective value greater than or equal to that of the optimum rank-1 solution. The elimination is achieved by penalizing the directions associated with the eigenvectors corresponding to the second-largest eigenvalues of previous high-rank solutions. Specifically, a quadratic matrix product penalty term of the form $\zeta\sum_{r}\mathbf{u}_r^H\mathbf{W}\mathbf{u}_r$ is applied to the per-AP power constraints, where $\zeta$ is a penalty factor and $\mathbf{u}_r$, $\forall r$, denote the eigenvectors associated with the second-largest eigenvalues from previous iterations with higher-rank solutions.

In this paper, we propose a new reformulation for the penalty term that is suitable for developing a computationally fast and effective ADMM algorithm. The penalty can be incorporated in the per-AP power constraints by updating the matrices $\mathbf{D}_l, \forall l$ for the subsequent SEA iterations as follows:
\begin{equation}
\begin{split}
    \textrm{tr}\left(\mathbf{D}_l\mathbf{W}\right) + \zeta\sum_{r}\mathbf{u}_r^H\mathbf{W}\mathbf{u}_r &= \textrm{tr}\left(\mathbf{D}_l\mathbf{W}\right) + \textrm{tr}\left(\zeta\sum_{r}\mathbf{U}_r\mathbf{W}\right)\\
    &= \textrm{tr}\left(\left(\mathbf{D}_l + \zeta\sum_{r}\mathbf{U}_r\right)\mathbf{W}\right)\\
    &= \Biggl\langle\mathbf{D}_l + \zeta\sum_{r}\mathbf{U}_r, \mathbf{W}\Biggr\rangle.
\end{split}
\end{equation}
That is,
\begin{equation}
    \mathbf{D}_l \leftarrow \mathbf{D}_l + \zeta\sum_{r}\mathbf{U}_r, \quad \forall l \in \{1, \hdots, L\},
\end{equation}
where $\mathbf{U}_r = \mathbf{u}_r\mathbf{u}_r^H$. Note that the penalty term can be incorporated with other constraints as well. Since $\mathbf{D}_l$, $\forall l$ are composed of binary entries and $\mathbf{U}_r$, $\forall r$ are built using unit-norm eigenvectors, the magnitude of the variables associated with the penalty parameter is fixed for fixed $L$ and $N$, and is mildly affected by changes in them. On the other hand, this penalty term is added onto the $L$ power constraints. For this reason, we choose to have $\zeta = 0.5/L$ in order not to introduce unnecessarily large penalty if $L$ becomes exceedingly large.

\subsection{Near-Global Optimality}

The global optimal solutions to the multicast MMF and QoS optimization problems are generally unattainable in an efficient manner, since both are non-convex and NP-hard problems. The state-of-the-art SCA technique begins with a feasible initial point and converges only to a local optimum, which may be significantly different from the global optimum. In contrast, the proposed SEA starts from an infeasible (higher-rank) solution and gradually moves toward the feasible region by incorporating penalty terms. Consequently, the resulting solution lies on the boundary of the feasible set, where the global optimal rank-1 solution to the original problem is known to exist. Moreover, the rank-1 solution obtained through SEA corresponds exactly to the global optimum of the non-convex penalized problems, where the optimality gap introduced by relaxation becomes $0$. Let us consider the QoS problem as an example, since the same approach and conclusions are applicable to both problems. By using $\mathbf{W} = \mathbf{ww}^H$ and reformulating the penalty term accordingly, the derived solution by the SEA is the global optimum to the penalized non-convex QoS problem which can be rewritten as follows:
\begin{subequations}
\begin{align}
    \mathop{\mathrm{minimize}}\limits_{\mathbf{w}} &\hspace{-0.5em}\max_{\hspace{0.5em} l \in \{1, \hdots, L\}} \frac{1}{P_{l,\mathrm{max}}}\Biggl(\norm{\mathbf{w}_{l}}_2^2 + \zeta\sum_{r}|\mathbf{u}_{r}^H\mathbf{w}|^2\Biggr) \\
    \textrm{s.t.} &\quad \frac{\left|\mathbf{h}_{k}^H\mathbf{w}\right|^2}{\sigma_{k}^2} \geq \gamma_k, \quad \forall k \in \{1, \hdots, K_g\}.
\end{align}%
\label{QoS_problem_reform}
\end{subequations}%

The $\mathbf{u}_r$, $\forall r$ in the additive penalty terms represent the eigenvectors corresponding to the second-largest eigenvalues of the higher-rank solutions obtained in previous SEA iterations. Since the solution matrix to the relaxed SDP is Hermitian positive semidefinite with distinct non-zero eigenvalues, its eigenvectors are mutually orthogonal. Moreover, the eigenvectors associated with the second-largest eigenvalues cannot produce an optimal rank-1 solution, since the dominant eigenvector of that solution will always yield a larger objective value. Consequently, penalizing these eigenvectors has minor influence on subsequent elimination iterations, which eventually drives the solution toward a near-global optimal rank-1 result. Numerical simulations showing the impact of these penalty terms on the optimality gap are provided in \cite{zaher2026cell}. It is worth mentioning that no theoretical global optimality guarantee is available for the multicast beamforming optimization problem due to its NP-hard nature, however, the near-global optimality claim for the SEA is rather supported by the small gap to the SDR bound, the SEA rank reduction mechanism and numerical evidence. 

\section{SEA-ADMM: Efficient Multicast Beamforming}\label{SEA_ADMM}

In this section, we develop a general SEA-ADMM framework to solve the MMF and QoS multicast beamforming problems. The proposed algorithm makes use of SDR, the SEA and ADMM to find near-global optimal rank-1 beamforming solutions to the NP-hard MMF and QoS problems in a cell-free massive MIMO network subject to per-AP power constraints.

\subsection{A Fast SEA-ADMM Algorithm For Multicast Beamforming}

The subsequent part will focus on the MMF problem \eqref{MMF_problem3}. Since the problem is convex and satisfies Slater's condition, strong duality holds and it is easier to solve the dual SDP problem.

\vspace{-0.5em}\begin{theorem}
    Define the adjoint operator of $\mathcal{H}$ as $\mathcal{H}^H\left(\cdot\right): \mathbb{R}^K \rightarrow \mathbb{H}^N$, which is given by $\mathcal{H}^H\left(\mathbf{y}\right) = \sum_{k = 1}^Ky_k\mathbf{H}_k$. Similarly, the adjoint operator of $\mathcal{D}$ is $\mathcal{D}^H\left(\cdot\right): \mathbb{R}^L \rightarrow \mathbb{H}^N$, and is given by $\mathcal{D}^H\left(\mathbf{z}\right) = \sum_{l = 1}^Lz_l\mathbf{D}_l$.
    The dual SDP problem to problem \eqref{MMF_problem3} is given by
    \begin{subequations}
    \begin{align}
        \min_{\mathbf{y}, \mathbf{z}, \mathbf{S}} \quad &\mathbf{z}^T\mathbf{p}\\
        \textrm{s.t.} \quad &\mathcal{H}^H\left(\mathbf{y}\right) + \mathbf{S} = \mathcal{D}^H\left(\mathbf{z}\right),\\
        & \norm{\mathbf{y}}_1 = 1,\quad \mathbf{y}\geq \mathbf{0},\quad \mathbf{z} \geq \mathbf{0},\\
        &\mathbf{S} \succeq \mathbf{0}.
    \end{align}
    \label{dualMMF}%
    \end{subequations}
    where $\mathbf{y} \in \mathbb{R}^K_+$ and $\mathbf{z} \in \mathbb{R}^L_+$ denote the nonnegative Lagrange multipliers of the constraints in \eqref{MMF_SNR} and \eqref{MMF_pow}, respectively, and $\mathbf{S} \in \mathbb{H}^N_+$ denotes the positive semidefinite Lagrange multiplier of the constraint in \eqref{MMF_psd}.
    \begin{IEEEproof}
    The proof is given in Appendix A.
    \end{IEEEproof}
\end{theorem}

We define the indicator functions $\mathbb{I}_{\Delta}\left(\mathbf{y}\right)$ and $\mathbb{I}_{\mathbf{z} \geq \mathbf{0}}\left(\mathbf{z}\right)$ as
\begin{equation}
    \mathbb{I}_{\Delta}\left(\mathbf{y}\right) =
    \begin{cases}
        0 &\textrm{for } \mathbf{y} \in \Delta, \\
        +\infty &\textrm{otherwise,}
    \end{cases}
    \label{indicator1}
\end{equation}
\begin{equation}
    \mathbb{I}_{\mathbf{z} \geq \mathbf{0}}\left(\mathbf{z}\right) =
    \begin{cases}
        0 &\textrm{for } \mathbf{z} \geq \mathbf{0}, \\
        +\infty &\textrm{otherwise,}
    \end{cases}
    \label{indicator2}
\end{equation}
where $\Delta$ represents the standard simplex defined as $\Delta = \{\mathbf{y}: \norm{\mathbf{y}}_1 = 1, \mathbf{y} \geq \mathbf{0}\}$.
Consequently, an equivalent reformulation of the dual SDP problem is given by
\begin{subequations}
\begin{align}
    \min_{\mathbf{y}, \mathbf{z}, \mathbf{S}} \quad &\mathbf{z}^T\mathbf{p} + \mathbb{I}_{\Delta}\left(\mathbf{y}\right) + \mathbb{I}_{\mathbf{z} \geq \mathbf{0}}\left(\mathbf{z}\right)\\
    \textrm{s.t.} \quad &\mathcal{H}^H\left(\mathbf{y}\right) + \mathbf{S} = \mathcal{D}^H\left(\mathbf{z}\right),\label{linear_constraint}\\
    &\mathbf{S} \succeq \mathbf{0}.
\end{align}
\label{dualMMF_reform}%
\end{subequations}
As such, the augmented Lagrangian, in scaled form, for the dual SDP corresponding to the linear constraints can be written as
\begin{equation}
\begin{split}
    L_{\rho}&\left(\mathbf{y},\mathbf{z},\mathbf{S},\overbar{\mathbf{W}}\right) = \mathbf{z}^T\mathbf{p} + \mathbb{I}_{\Delta}\left(\mathbf{y}\right) + \mathbb{I}_{\mathbf{z} \geq \mathbf{0}}\left(\mathbf{z}\right) \\
    &+ \frac{\rho}{2}\norm{\mathcal{H}^H\left(\mathbf{y}\right) + \mathbf{S} - \mathcal{D}^H\left(\mathbf{z}\right) + \overbar{\mathbf{W}}}_F^2,
\end{split}
\end{equation}
where $\overbar{\mathbf{W}} = \mathbf{W}/\rho$ is the scaled dual variable and $\rho \geq 0$ is the penalty parameter associated with the linear constraints in \eqref{linear_constraint}.

Utilizing the ADMM algorithm \cite{boyd2011distributed}, the minimization is done with respect to the blocks of variables $\{\mathbf{y}, \mathbf{z}\}$ and $\mathbf{S}$ separately while the other variables are kept fixed. At each iteration, the following updates are computed sequentially:
\begin{align}
    &\left\{\mathbf{y}^{i+1}, \mathbf{z}^{i+1}\right\} = \mathrm{arg\,}\mathop{\mathrm{min}}\limits_{\mathbf{y}, \mathbf{z}}\,L_{\rho}\left(\mathbf{y},\mathbf{z},\mathbf{S}^i,\overbar{\mathbf{W}}^i\right),\label{yz_update}\\
    &\mathbf{S}^{i+1} = \mathrm{arg\,}\mathop{\mathrm{min}}\limits_{\mathbf{S}}\,L_{\rho}\left(\mathbf{y}^{i+1},\mathbf{z}^{i+1},\mathbf{S},\overbar{\mathbf{W}}^i\right), \textrm{ s.t. } \mathbf{S} \succeq \mathbf{0}, \label{S_update}\\
    &\overbar{\mathbf{W}}^{i+1} = \overbar{\mathbf{W}}^i + \mathcal{H}^H\left(\mathbf{y}^{i+1}\right) + \mathbf{S}^{i+1} - \mathcal{D}^H\left(\mathbf{z}^{i+1}\right).
    \label{W_update}
\end{align}

\subsubsection{$\{\mathbf{y}, \mathbf{z}\}$-Update}

The update of the first block of variables $\{\mathbf{y}, \mathbf{z}\}$ in \eqref{yz_update} is equivalent to solving the following problem:
\begin{subequations}
\begin{align}
    \min_{\mathbf{y},\mathbf{z}} \quad &\mathbf{z}^T\mathbf{p} + \frac{\rho}{2}\norm{\mathcal{H}^H\left(\mathbf{y}\right) + \mathbf{S}^i - \mathcal{D}^H\left(\mathbf{z}\right) + \overbar{\mathbf{W}}^i}_F^2\\
    \textrm{s.t.} \quad & \mathbf{y} \in \Delta, \quad \mathbf{z} \geq \mathbf{0}.
\end{align}
\label{yz_update_problem}%
\end{subequations}

\vspace{-1.5em}\begin{theorem}
    Let $\textrm{vec}(\mathbf{A})$ denote the column-wise vectorization of the matrix $\mathbf{A}$. Define the mapping matrices $\mathbf{H} \in \mathbb{C}^{K \times (LN)^2}$ and $\mathbf{D} \in \mathbb{C}^{L \times (LN)^2}$, where $\mathbf{H} = [\textrm{vec}(\mathbf{H}_1)^H; \hdots; \textrm{vec}(\mathbf{H}_K)^H]$ and $\mathbf{D} = [\textrm{vec}(\mathbf{D}_1)^H; \hdots; \textrm{vec}(\mathbf{D}_L)^H]$. Moreover, define $\mathbf{r} = \textrm{vec}(\mathbf{S}^i + \overbar{\mathbf{W}}^i)$ and let $\mathbf{x} = [\mathbf{y}^T, \mathbf{z}^T]^T \in \mathbb{R}^{K+L}$. The update of the $\{\mathbf{y}, \mathbf{z}\}$ variable block in problem \eqref{yz_update_problem} can be reformulated as the following quadratic program (QP) in standard form:
    \begin{subequations}
    \begin{align}
        \min_{\mathbf{x}} \quad &\frac{1}{2}\mathbf{x}^T\mathbf{Qx} + \mathbf{c}^T\mathbf{x}\\
        \textrm{s.t.} \quad &\mathbf{x} \in \mathcal{C},
    \end{align}
    \label{QP}%
    \end{subequations}
    where 
    \[
    \mathbf{Q} = \rho
    \begin{bmatrix}
    \hspace{7pt}\mathbf{HH}^H & -\mathbf{HD}^H\\
    -\mathbf{DH}^H & \hspace{7pt}\mathbf{DD}^H
    \end{bmatrix}
    ,\quad
    \mathbf{c} = 
    \begin{bmatrix}
    \rho\hspace{1pt}\Re\{\mathbf{Hr}\}\\
    \mathbf{p} - \rho\hspace{1pt}\Re\{\mathbf{Dr}\}
    \end{bmatrix}
    ,
    \]
    and $\mathcal{C} = \{[\mathbf{y}^T, \mathbf{z}^T]^T: \mathbf{
    y} \in \Delta, \mathbf{z} \geq \mathbf{0}\}$.
\begin{IEEEproof}
    The proof is given in Appendix B.
    \end{IEEEproof}
\end{theorem}

By introducing an auxiliary variable $\mathbf{v} \in \mathbb{R}^{K+L}$ and the indicator function 
\begin{equation}
    \mathbb{I}_{\mathcal{C}}(\mathbf{v}) = 
    \begin{cases}
        0 &\textrm{for } \mathbf{v} \in \mathcal{C}, \\
        +\infty &\textrm{otherwise,}
    \end{cases}
\end{equation}
problem \eqref{QP} can be reformulated as
\begin{subequations}
\begin{align}
    \min_{\mathbf{x}} \quad &\frac{1}{2}\mathbf{x}^T\mathbf{Qx} + \mathbf{c}^T\mathbf{x} + \mathbb{I}_{\mathcal{C}}(\mathbf{v})\\
    \textrm{s.t.} \quad &\mathbf{x} = \mathbf{v}.\label{equality_constraints}
\end{align}
\label{QP_reform}%
\end{subequations}
This problem can be efficiently solved by an inner ADMM algorithm to find the optimal $\{\mathbf{y}, \mathbf{z}\}$-update. We denote the scaled dual variable associated with the equality constraints \eqref{equality_constraints} by $\bar{\mathbf{t}} \in \mathbb{R}^{K+L}$. Since the variable $\mathbf{x}$ comprises $\mathbf{y}$ and $\mathbf{z}$, which represent the dual variables associated with the SNR and power constraints and have different orders-of-magnitude, the penalties associated with the equality constraints \eqref{equality_constraints} should not have the same value. Accordingly, the augmented Lagrangian of problem \eqref{QP_reform} is formulated as
\begin{equation}
\begin{split}
    L_\mu\left(\mathbf{x}, \mathbf{v}, \bar{\mathbf{t}}\right) &= \frac{1}{2}\mathbf{x}^T\mathbf{Qx} + \mathbf{c}^T\mathbf{x} + \mathbb{I}_{\mathcal{C}}(\mathbf{v})\\
    & \quad +\frac{1}{2} \left(\mathbf
    x - \mathbf{v} + \bar{\mathbf{t}}\right)^T\mathbf{R}\left(\mathbf
    x - \mathbf{v} + \bar{\mathbf{t}}\right),
\end{split}
\label{lagrangian2}%
\end{equation}
where $\mathbf{R} \in \mathbb{R}^{(K+L)\times(K+L)}$ is a diagonal matrix, such that the diagonal elements, denoted as $\textrm{diag}(\mathbf{R})$, comprise the positive penalties associated with the equality constraints \eqref{equality_constraints} and are given by $\textrm{diag}(\mathbf{R}) = [\mu_1^s, \hdots, \mu_K^s, \mu_1^p, \hdots, \mu_L^p]^T$.

Utilizing the ADMM algorithm, the update of the variable blocks $\mathbf{x}$ and $\mathbf{v}$ is done successively, while the other variables are kept fixed. Specifically, the following updates are computed at every inner ADMM iteration:
\begin{align}
    &\mathbf{x}^{j+1} = \mathrm{arg\,}\mathop{\mathrm{min}}\limits_{\mathbf{x}}\,L_{\mu}\left(\mathbf{x},\mathbf{v}^j,\bar{\mathbf{t}}^j\right),\\
    &\mathbf{v}^{j+1} = \mathrm{arg\,}\mathop{\mathrm{min}}\limits_{\mathbf{v}}\,L_{\mu}\left(\mathbf{x}^{j+1},\mathbf{v},\bar{\mathbf{t}}^j\right),\label{v_update}\\
    &\bar{\mathbf{t}}^{j+1} = \bar{\mathbf{t}}^j + \mathbf{x}^{j+1} - \mathbf{v}^{j+1}. \label{t_update}
\end{align}

The minimization of \eqref{lagrangian2} with respect to $\mathbf{x}$ is done by equating its first-order derivative to zero, that is
\begin{equation}
    \mathbf{Qx}^{j+1} + \mathbf{c} + \mathbf{R}\left(\mathbf{x}^{j+1} - \mathbf{v} + \bar{\mathbf{t}}\right) = 0.
\end{equation}
Hence, the closed-form $\mathbf{x}$-update is given by
\begin{equation}
    \mathbf{x}^{j+1} = \left(\mathbf{Q + R}\right)^{-1}\left(-\mathbf{c} + \mathbf{R}\left(\mathbf{v} - \bar{\mathbf{t}}\right)\right).
    \label{x_update}
\end{equation}
Note that the matrices $\mathbf{Q}$ and $\mathbf{R}$ are independent of both the inner and outer ADMM updates. Accordingly, the matrix inverse in \eqref{x_update} needs to be computed only once before the start of the algorithm. Further, the vector $\mathbf{c}$ is independent of the inner ADMM updates, and so is required to be computed once for each outer ADMM update. As a result, a fast inner ADMM update is attained since the remaining term is cheap to compute.

The update of the auxiliary variable $\mathbf{v}$ in problem \eqref{v_update} is equivalent to solving the following problem:
\begin{subequations}
\begin{align}
    \min_{\mathbf{v}} \quad &\norm{\mathbf{v} - \bigl(\mathbf{x}^{j+1} + \bar{\mathbf{t}}^j\bigr)}_2^2 \\
    \textrm{s.t.} \quad &\mathbf{v} \in \mathcal{C}.
\end{align}
\end{subequations}

Let $\mathbf{v}' = \mathbf{x}^{j+1} + \bar{\mathbf{t}}^j$. By splitting $\mathbf{v}'$ into $\mathbf{y}' \in \mathbb{R}^K$ and $\mathbf{z}' \in \mathbb{R}^L$, the optimal inner $\{\mathbf{y}, \mathbf{z}\}$-update can be written as
\begin{align}
    &\mathbf{y}^{j+1} = \boldsymbol{\Pi}_{\Delta}\left(\mathbf{y}'\right),\\
    &\mathbf{z}^{j+1} = \mathrm{max}\left(\mathbf{z}', \mathbf{0}\right),
\end{align}
where $\boldsymbol{\Pi}_{\Delta}(\mathbf{y}')$ represents the projection of $\mathbf{y}'$ onto the standard simplex $\Delta$. This operation can be done efficiently using a sorting-based algorithm as in \cite{duchi2008efficient}. The algorithm is provided in Appendix C for completeness. The optimal $\mathbf{v}$-update is then $\mathbf{v}^{j+1} = [(\mathbf{y}^{j+1})^T, (\mathbf{z}^{j+1})^T]^T$. Note that the iteration index for the outer ADMM algorithm is omitted for the variables $\{\mathbf{y}, \mathbf{z}\}$ to simplify the notation.

\subsubsection{$\mathbf{S}$-Update}

The $\mathbf{S}$-update in problem \eqref{S_update} can be reformulated as
\begin{subequations}
    \begin{align}
        \min_{\mathbf{S}} \quad &\norm{\mathbf{S} - \mathbf{X}^{i+1}}_F^2 \\
    \textrm{s.t.} \quad &\mathbf{S} \succeq \mathbf{0},
    \end{align}
\end{subequations}
where $\mathbf{X}^{i+1} = \mathcal{D}^H\left(\mathbf{z}^{i+1}\right) - \mathcal{H}^H\left(\mathbf{y}^{i+1}\right) - \overbar{\mathbf{W}}^i$. The $\mathbf{S}$-update is then the projection of $\mathbf{X}^{i+1}$ onto the positive semidefinite cone, and is given by
\begin{equation}
    \mathbf{S}^{i+1} = \mathbf{X}_+^{i+1} \triangleq \mathbf{Q}_+^{i+1}\boldsymbol{\Sigma}_+^{i+1}(\mathbf{Q}_+^{i+1})^H,
\end{equation}
where $\boldsymbol{\Sigma}_+$ is a diagonal matrix with the non-negative eigenvalues of $\mathbf{X}^{i+1}$ and $\mathbf{Q}_+^{i+1}$ denotes a matrix with the corresponding eigenvectors as columns.

\subsection{QoS Multicast Beamforming}\label{QoS_subsection}

In this section, we will devise another variant of our SEA-ADMM algorithm adapted to the QoS multicast problem. By examining the problem formulations for the MMF multicast problem \eqref{MMF_problem3} and the QoS multicast problem \eqref{QoS_problem2}, it is clear that the same approach for solving the MMF problem in the previous section can be adopted to devise the solution to the QoS problem. The same dual variables corresponding to the linear and SDP constraints will be used to showcase similarities and differences between the two problems.

To construct the dual SDP for the QoS problem, the Lagrangian for the QoS problem \eqref{QoS_problem2} is formulated as
\begin{equation}
\begin{split}
    L_{\textrm{QoS}}\left(\mathbf{W}, x, \mathbf{y}, \mathbf{z}, \mathbf{S}\right) &= x - \langle \mathbf{y}, \mathcal{H}\left(\mathbf{W}\right) - \boldsymbol{\gamma}\rangle  \\
    &+ \langle\mathbf{z}, \mathcal{D}\left(\mathbf{W}\right)  - \mathbf{p}x\rangle - \langle\mathbf{S}, \mathbf{W}\rangle.
\end{split}
\label{QoS_lag}
\end{equation}

Analogous to the derivation of the dual SDP problem to the MMF problem in Theorem 1, by equating the derivatives of the Lagrangian with respect to $\mathbf{W}$ and $x$ to $\mathbf{0}$, the dual SDP for the QoS problem can be formulated as
\begin{subequations}
\begin{align}
    \min_{\mathbf{y}, \mathbf{z}, \mathbf{S}} \quad &-\mathbf{y}^T\boldsymbol{\gamma}\\
    \textrm{s.t.} \quad &\mathcal{H}^H\left(\mathbf{y}\right) + \mathbf{S} = \mathcal{D}^H\left(\mathbf{z}\right),\\
    & \mathbf{z}^T\mathbf{p} = 1,\quad \mathbf{y}\geq \mathbf{0},\quad \mathbf{z} \geq \mathbf{0},\\
    &\mathbf{S} \succeq \mathbf{0}.
\end{align}
\label{dualQoS}%
\end{subequations}

Similar to \eqref{indicator1} and \eqref{indicator2}, we define the indicator functions $\mathbb{I}_{\Delta'}\left(\mathbf{z}\right)$ and $\mathbb{I}_{\mathbf{y} \geq \mathbf{0}}\left(\mathbf{y}\right)$,
where $\Delta'$ represents the simplex defined as $\Delta' = \{\mathbf{z}: \mathbf{z}^T\mathbf{p} = 1, \mathbf{z} \geq \mathbf{0}\}$.
Hence, an equivalent reformulation of the dual SDP problem in this case is given by
\begin{subequations}
\begin{align}
    \min_{\mathbf{y}, \mathbf{z}, \mathbf{S}} \quad &\hspace{-2pt}-\mathbf{y}^T\boldsymbol{\gamma} + \mathbb{I}_{\Delta'}\left(\mathbf{z}\right) + \mathbb{I}_{\mathbf{y} \geq \mathbf{0}}\left(\mathbf{y}\right)\\
    \textrm{s.t.} \quad &\mathcal{H}^H\left(\mathbf{y}\right) + \mathbf{S} = \mathcal{D}^H\left(\mathbf{z}\right),\\
    &\mathbf{S} \succeq \mathbf{0}.
\end{align}
\label{dualQoS_reform}%
\end{subequations}
Consequently, the augmented Lagrangian in scaled form for the dual SDP corresponding to the linear constraints is formulated as
\begin{equation}
\begin{split}
    L_{\rho}&\left(\mathbf{y},\mathbf{z},\mathbf{S},\overbar{\mathbf{W}}\right) = -\mathbf{y}^T\boldsymbol{\gamma} + \mathbb{I}_{\Delta'}\left(\mathbf{z}\right) + \mathbb{I}_{\mathbf{y} \geq \mathbf{0}}\left(\mathbf{y}\right) \\
    &+ \frac{\rho}{2}\norm{\mathcal{H}^H\left(\mathbf{y}\right) + \mathbf{S} - \mathcal{D}^H\left(\mathbf{z}\right) + \overbar{\mathbf{W}}}_F^2,
\end{split}
\end{equation}

Similar to the case of the MMF problem, the minimization is done using the ADMM algorithm \cite{boyd2011distributed} with respect to the blocks of variables $\{\mathbf{y}, \mathbf{z}\}$ and $\mathbf{S}$ separately, while the other variables are kept fixed. The variables' update is computed as given in \eqref{yz_update}-\eqref{W_update}.

For the update of the first block of variables $\{\mathbf{y}, \mathbf{z}\}$, the following equivalent problem is solved:
\begin{subequations}
\begin{align}
    \min_{\mathbf{y},\mathbf{z}} \quad &\hspace{-2pt}-\mathbf{y}^T\boldsymbol{\gamma} + \frac{\rho}{2}\norm{\mathcal{H}^H\left(\mathbf{y}\right) + \mathbf{S}^i - \mathcal{D}^H\left(\mathbf{z}\right) + \overbar{\mathbf{W}}^i}_F^2\\
    \textrm{s.t.} \quad & \mathbf{z} \in \Delta', \quad \mathbf{y} \geq \mathbf{0}.
\end{align}
\label{yz_update_QoS}
\end{subequations}
This problem can be reformulated as the following QP in standard form:
\begin{subequations}
\begin{align}
    \min_{\mathbf{x}} \quad &\frac{1}{2}\mathbf{x}^T\mathbf{Q}'\mathbf{x} + \mathbf{c}'^T\mathbf{x}\\
    \textrm{s.t.} \quad &\mathbf{x} \in \mathcal{C}',
\end{align}
\label{QP2}%
\end{subequations}
where 
\[
\mathbf{Q}' = \rho
\begin{bmatrix}
\hspace{7pt}\mathbf{HH}^H & -\mathbf{HD}^H\\
-\mathbf{DH}^H & \hspace{7pt}\mathbf{DD}^H
\end{bmatrix}
,\quad
\mathbf{c}' = 
\begin{bmatrix}
\rho\hspace{1pt}\Re\{\mathbf{Hr}\} - \boldsymbol{\gamma}\\
-\rho\hspace{1pt}\Re\{\mathbf{Dr}\}
\end{bmatrix}
,
\]
and $\mathcal{C}' = \{[\mathbf{y}^T, \mathbf{z}^T]^T: \mathbf{
z} \in \Delta', \mathbf{y} \geq \mathbf{0}\}$. The proof follows a similar procedure to that of Theorem 2 given in Appendix B and thus is omitted here. Problem \eqref{QP2} can be efficiently solved using an inner ADMM algorithm to find the optimal $\{\mathbf{y}, \mathbf{z}\}$-update, as detailed in the previous subsection for the MMF problem. In addition, the update of the second block of variables $\mathbf{S}$ and the dual variables $\overbar{\mathbf{W}}$ are the same as in the case of the MMF problem.

\subsection{Convergence Criteria}

For the outer ADMM updates, the stopping criteria are selected to verify the convergence of primal and dual variables. The ADMM updates terminate when the following two conditions are satisfied at the $i^{th}$ iteration:
\begin{align}
    \frac{\bigl|\textrm{tr}\left(\overbar{\mathbf{W}}^i - \overbar{\mathbf{W}}^{i-1}\right)\bigr|}{\textrm{tr}\left(\overbar{\mathbf{W}}^i\right)} &< \epsilon_{\mathrm{dual}}, \label{W_cond}\\
    \frac{\norm{\mathbf{S}^i - \mathbf{S}^{i-1}}_F}{\norm{\mathbf{S}^i}_F} &< \epsilon_{\mathrm{prim}}, \label{S_cond}
\end{align}
where $\epsilon_\mathrm{dual} > 0$ and $\epsilon_\mathrm{prim} > 0$ are predefined stopping conditions. The convergence of the ADMM algorithm is guaranteed by well-known ADMM convergence results for convex problems \cite{boyd2011distributed}. We highlight that the stopping criterion for the dual variable $\overbar{\mathbf{W}}$ directly translates into the relative change in the total transmit power of the precoder for the original optimization problem. To avoid early termination due to stagnation, both conditions are deemed necessary to achieve the best possible outcome. As for the inner ADMM algorithm with cheap updates, we avoid computing stopping criteria to reduce the complexity per inner iteration. Instead, the algorithm is run for a fixed number of iterations $T$.

The SEA-ADMM algorithm for MMF and QoS multicast beamforming optimization is summarized in Algorithm \ref{alg1}. Note that $P_T$ represents the sum of the maximum transmit powers of all APs.

\begin{algorithm}
        \caption{Multicast Beamforming via SEA-ADMM}
        \label{alg1}
        \noindent\textbf{Input:} ADMM penalty parameters $\rho$ and $\mathbf{R}$, and stopping conditions $\epsilon_\mathrm{dual} > 0$ and $\epsilon_\mathrm{prim} > 0$. Number of inner ADMM iterations $T$. Initialize $\mathbf{y}^0 = \mathbf{z}^0 = \bar{\mathbf{t}}^0 = \mathbf{0}$, $\mathbf{S}^0 = \mathbf{0}_{LN \times LN}$, $ \overbar{\mathbf{W}}^0 = \frac{P_T}{\rho LN}\mathbf{I}_{LN}$.
        
	\begin{algorithmic}[1]
            \REPEAT \label{ADMM_start}
            \STATE Update the first block of variables $\{\mathbf{y}, \mathbf{z}\}$
            \begin{equation}
                \left\{\mathbf{y}^{i+1}, \mathbf{z}^{i+1}\right\} = \mathrm{arg\,}\mathop{\mathrm{min}}\limits_{\mathbf{y}, \mathbf{z}}\,L_{\rho}\left(\mathbf{y},\mathbf{z},\mathbf{S}^i,\overbar{\mathbf{W}}^i\right).
            \end{equation}
            \STATE Update the second block of variables $\mathbf{S}$
            \begin{equation}
                \mathbf{S}^{i+1} = \mathrm{arg\,}\mathop{\mathrm{min}}\limits_{\mathbf{S}}\,L_{\rho}\left(\mathbf{y}^{i+1},\mathbf{z}^{i+1},\mathbf{S},\overbar{\mathbf{W}}^i\right), \textrm{ s.t. } \mathbf{S} \succeq \mathbf{0}.
            \end{equation}
            \STATE Update the dual variables $\overbar{\mathbf{W}}$
            \begin{equation}
                \overbar{\mathbf{W}}^{i+1} = \overbar{\mathbf{W}}^i + \mathcal{H}^H\left(\mathbf{y}^{i+1}\right) + \mathbf{S}^{i+1} - \mathcal{D}^H\left(\mathbf{z}^{i+1}\right).
            \end{equation}
            \UNTIL Convergence criteria in \eqref{W_cond} and \eqref{S_cond} are met. \label{ADMM_end}
            \STATE Set $r \leftarrow 1$.
            \WHILE{$\textrm{rank}\bigl(\overbar{\mathbf{W}}\bigr) \neq 1$} \label{SEA_start}
            \STATE Set $\mathbf{U}_r \leftarrow \mathbf{u}_r\mathbf{u}_r^H$, where $\mathbf{u}_r$ is the second strongest eigenvector of $\overbar{\mathbf{W}}$.
            \STATE Update $\mathbf{D}_l \leftarrow \mathbf{D}_l + \zeta\mathbf{U}_r$, $\forall l \in \{1, \hdots, L\}$.
            \STATE Repeat steps $\ref{ADMM_start}$-$\ref{ADMM_end}$.
            \STATE Set $r \leftarrow r + 1$.
            \ENDWHILE \label{SEA_end}
            \STATE Compute $\mathbf{W} = \rho\overbar{\mathbf{W}}$.
\end{algorithmic}
\textbf{Output:} The near-optimal rank-$1$ solution.
\end{algorithm}

Unlike methods based on the SCA technique (the prior state-of-the-art in multicast beamforming optimization), a key benefit of the proposed algorithm is that it does not require a feasible initial solution to the original problems. This is shown in the initialization utilized in Algorithm \ref{alg1}. Finding a suitable low-complexity initialization can be particularly challenging for SCA-based algorithms as the total number of antennas and UEs increases. Examples of initialization algorithms and their effect on the convergence speed of SCA-based algorithms can be found in \cite{zhang2023ultra}.

\subsection{Complexity Analysis}

The most computationally demanding operations for the proposed SEA-ADMM algorithm lie in computing the matrix $\mathbf{Q}$ ($\mathbf{Q}'$ for the QoS problem) and its inverse in \eqref{x_update}, which has a complexity of order $\mathcal{O}((L + K)^2(LN)^2$ and $\mathcal{O}((L + K)^3)$, respectively. This operation needs to be computed only once for every outer SEA iteration, and can be reused in subsequent ADMM iterations. The computation of $\mathbf{c}$ ($\mathbf{c}'$ for the QoS problem) requires $\mathcal{O}((L + K)(LN)^2)$ operations, and needs to be performed once for every outer ADMM iteration. The dominant component in terms of complexity for the inner ADMM updates lies in the projection to the simplex, which requires $\mathcal{O}(K\,\textrm{log}_2(K))$ ($\mathcal{O}(L\,\textrm{log}_2(L))$ for the QoS problem) operations.

For a general-purpose solver that employs interior-point methods, solving the relaxed MMF or QoS problems requires a worst-case complexity of $\mathcal{O}((LN)^{6.5} + (L + K)(LN)^{2.5})$ \cite{karipidis2008quality,zaher2026cell}. However, actual runtimes grow far slower with $LN$ than this worst-case bound, with a typical complexity per SEA iteration of order $\mathcal{O}((LN)^{3.5})$ to $\mathcal{O}((LN)^{4.5})$ depending on the problem structure \cite{karipidis2008quality,luo2010semidefinite}. The required number of SEA iterations for all simulated scenarios listed in the next section is generally between $1$-$10$, and is roughly the same when using ADMM or interior-point methods.

In Section \ref{results}, we will analyze the complexity and convergence behavior of the proposed SEA-ADMM algorithm numerically and compare it to relevant designs in the literature.

\section{Numerical Evaluation} \label{results}

In this section, we use Monte Carlo simulations to demonstrate the effectiveness of the proposed SEA-ADMM optimization procedure in solving the MMF and QoS multicast problems. We consider a cell-free massive MIMO network with $L = 9$ APs deployed on a square grid, serving an area of $750\,\textrm{m} \times 750\,\textrm{m}$. Each AP is equipped with a half-wavelength-spaced uniform linear array of $N = 4$ antennas, unless otherwise indicated. A wrap-around topology is employed to mitigate boundary effects. We assume $K \in \{10, 20, 30\}$ UEs, that are randomly and uniformly distributed within the area of interest, are jointly served by the APs using a single multicast transmission. In the case of the MMF problem, the maximum per-AP transmit power is set to $P_{l,\mathrm{max}} = 1$\,W, $\forall l$. As for the QoS problem, since all UEs in a multicast transmission need to successfully decode the transmitted data, a common SNR target $\gamma_c$ is assumed for all UEs, i.e., $\gamma_k = \gamma_c$, $\forall k$. Nonetheless, it is important to note that the proposed algorithm can also accommodate different SNR targets. The simulation parameters are summarized in Table \ref{params}. The channel between AP $l$ and an arbitrary UE $k$ is modeled by correlated Rayleigh fading as $\mathbf{h}_{kl} \sim \mathcal{N}_{\mathbb{C}}(\mathbf{0}, \mathbf{R}_{kl})$, where $\mathbf{R}_{kl} \in \mathbb{C}^{N \times N}$ represents the spatial correlation matrix, generated using the local scattering model in \cite{bjornson2017book}. The average channel gain, $\beta_{kl} = \frac{1}{N} \hspace{1pt}\textrm{tr}\hspace{-1pt}\left(\mathbf{R}_{kl}\right)$, is calculated using the 3GPP Urban Microcell model with correlated shadowing among the UEs. More precisely, the average channel gains are given by
\begin{equation}
\beta_{kl} = -30.5 - 36.7 \textrm{log}_{10}\left(\frac{d_{kl}}{1\,\textrm{m}}\right) + F_{kl} \hspace{2pt}\textrm{dB},
\label{pathloss}
\end{equation}
where $d_{kl}$ represents the distance between AP $l$ and UE $k$, and $F_{kl} \sim \mathcal{N}\left(0, 4^2\right)$ denotes the shadow fading. The shadowing is correlated between a given AP and different UEs as
\begin{equation}
\mathbb{E}\{F_{kl}F_{in}\} = 
\begin{cases}
    4^22^{-\delta_{ki}/9\,\textrm{m}}, & \textrm{for } l = n, \\
    0 & \textrm{for } l \neq n,
\end{cases}
\label{shadowing}
\end{equation}
where $\delta_{ki}$ is the distance between UE $k$ and UE $i$. Note that the correlation of shadowing between different APs, corresponding to the second case in \eqref{shadowing}, can be considered negligible due to the much larger distances between APs compared to those between UEs.

\begin{table}
\vspace{0.1cm}
\begin{center}
\caption{Cell-free network simulation parameters.}
\begin{tabular}{ |c|c| }
\hline
Area of interest & $750\,\textrm{m} \times 750\,\textrm{m}$ \\
Bandwidth & $20$\,MHz \\
Number of APs & $L = 9$ \\
Number of AP antennas & $N = 4$ \\
Number of UEs & $K = \left\{10, 20, 30\right\}$ \\
Per-AP transmit power (MMF) & $P_{l,\mathrm{max}} = 1$\,W \\
Common SNR target (QoS) & $\gamma_c = 255$ \\
Pathloss exponent & $\alpha = 3.67$ \\
DL noise power & $-94$\,dBm \\
\hline
\end{tabular}
\label{params}
\end{center}
\vspace{-1em}
\end{table} 

\subsection{Performance Analysis}

The SEA-ADMM penalty parameter for both the MMF and QoS multicast problems is set to $\rho = 0.2$, the inner ADMM penalty parameters corresponding to the $K$ constraints on the dual variable $\mathbf{y}$ are set to $\mu_k^s = 5 \times 10^6, \forall k$ and $\mu_k^s = 3 \times 10^6, \forall k$ for the MMF and QoS problems, respectively, whereas the penalties for the $L$ constraints on $\mathbf{z}$ are set to $\mu_l^p = 5, \forall l$ in both cases. The stopping conditions for the SEA-ADMM algorithm are set to $\epsilon_\mathrm{dual} = 2 \times 10^{-5}$ and $\epsilon_\mathrm{prim} = 7 \times 10^{-5}$, whereas the maximum number of iterations is set to $1000$.\footnote{The general guideline is to select penalty parameters that provide a good balance between the magnitude of the optimization objective and the introduced penalty term in order to satisfy its associated constraints without reducing the convergence speed toward the optimal objective. The stopping conditions and inner ADMM iterations reflect the desired/acceptable accuracy for the outer and inner ADMM algorithms, respectively. We have studied different values for the penalty parameters and stopping conditions and found that this combination provides good performance and convergence speed.} For every outer ADMM update, the inner ADMM updates are run for a fixed number of iterations $T = 50$. We use the same platform for performing the simulations, a 4-core Intel(R) Core i5-10310U CPU with 1.7 GHz base frequency and 4.4 GHz turbo frequency. All programs are written in Matlab. All curves are generated using $3000$ simulation samples, each incorporating different UE locations and channel realizations.

\begin{figure}
\centering
\setlength{\abovecaptionskip}{0.33cm plus 0pt minus 0pt}
\includegraphics[scale=0.465]{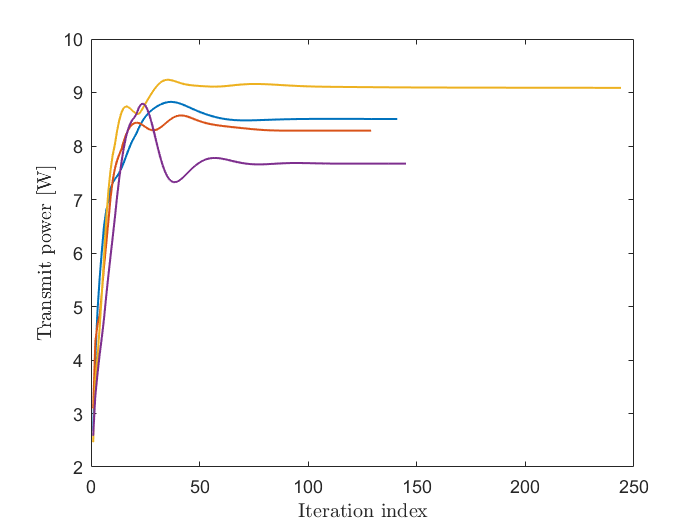}
\caption{Convergence behavior for the QoS problem: Transmit power vs. the outer ADMM iterations for the first SEA iteration, $K = 30$. The curves represent $4$ different realizations.}
\label{convergence}
\vspace{-1.1em}
\end{figure}

Fig. \ref{convergence} shows the convergence behavior of the transmit power for $4$ different simulation samples against the outer ADMM iterations for the first SEA iteration, which represents the SDR lower bound of the QoS problem. It can be seen that the first $20\sim30$ iterations expand the transmit power while tuning the precoder toward the optimal beamforming directions to satisfy the QoS constraints. Afterwards, fine-tuning of the precoder takes shape with a few oscillations of the transmit power, where the algorithm balances between minimizing the per-AP power and satisfying the QoS constraints. We have observed that typically $100\sim250$ outer ADMM iterations are required to converge for the simulated setups. The MMF problem exhibits similar convergence behavior.

\begin{figure}
\centering
\setlength{\abovecaptionskip}{0.33cm plus 0pt minus 0pt}
\includegraphics[scale=0.465]{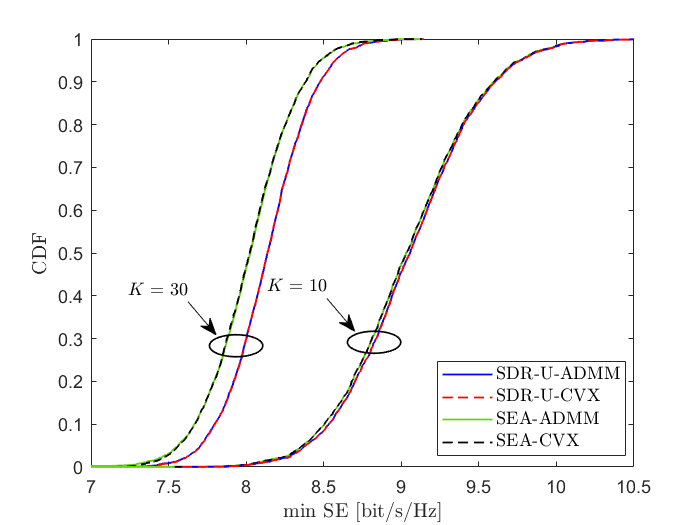}
\caption{CDF of the minimum SE for different numbers of UEs.}
\label{MMF_cdf1}
\end{figure}

\begin{figure}
\centering
\setlength{\abovecaptionskip}{0.33cm plus 0pt minus 0pt}
\includegraphics[scale=0.465]{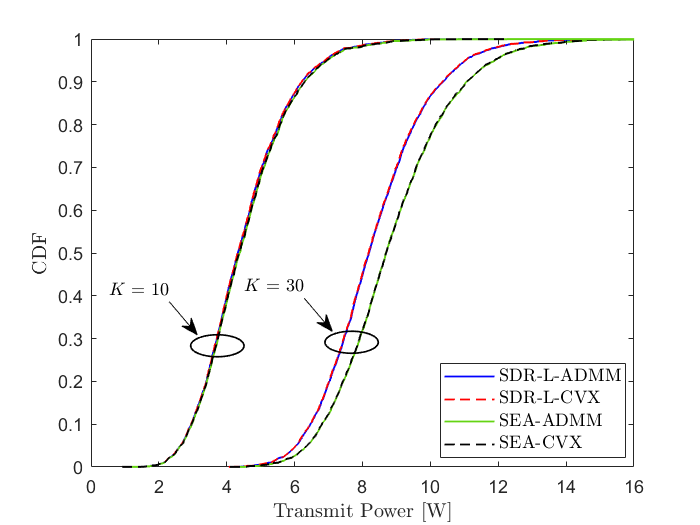}
\caption{CDF of the total transmit power for different numbers of UEs.}
\label{QoS_cdf1}
\end{figure}

Figs. \ref{MMF_cdf1} and \ref{QoS_cdf1} plot the cumulative distribution function (CDF) of the minimum SE (MMF) and the total transmit power (QoS) for the proposed SEA-ADMM algorithm for different numbers of UEs, respectively. The SEA with CVX \cite{cvx} utilized to solve the optimization problems is shown for comparison.
In Fig. \ref{MMF_cdf1}, the higher-rank SDR upper bound on the minimum SE utilizing the proposed ADMM- or CVX-based solution is provided. Similarly, the lower bound on the transmit power is shown in  Fig. \ref{QoS_cdf1}. It is clear that the proposed low-complexity SEA-ADMM algorithm is able to achieve the same performance for both the MMF and QoS multicast problems as its CVX-based counterpart, throughout all the simulated scenarios. For the case of $K = 10$ UEs, it can be seen that the gap between the higher-rank upper/lower bound is negligible since the output rank from solving the relaxed MMF and QoS problems is nearly 1. As the number of UEs increases, the rank of the solution matrix to the relaxed problems increases, requiring more iterations of the SEA and introducing a larger difference to the higher-rank solution. We highlight that this gap does not reflect the difference in performance between the proposed SEA-ADMM algorithm and the global optimum rank-1 solution, but it can only be viewed as an upper bound on the optimality gap.

To the best of our knowledge, the proposed SEA-ADMM algorithm is the first ADMM-based algorithm tailored for the MMF and QoS multicast beamforming optimization problems considering a cell-free massive MIMO network setting. The unique features of this setup are the multi-antenna APs and per-AP power budgets.
We emphasize that our previously proposed SEA-CVX method is used as a benchmark, as it has been established that it outperforms SCA-based methods when standard solvers are employed to handle the optimization problems \cite{zaher2026cell}. The superior performance over SCA-based techniques arises from their use of a gradient descent approach, which guarantees convergence only to a stationary point of the non-convex MMF and QoS multicast problems. In contrast, the proposed algorithm starts from the optimal higher-rank solution obtained via SDR and progressively reduces its rank by penalizing the eigenvectors corresponding to the second-largest eigenvalues. This iterative process continues until a near-optimal rank-1 solution is achieved within the orthogonal subspace of these eigenvectors. Throughout the iterations, the solution matrices remain Hermitian positive semidefinite with distinct non-zero eigenvalues, ensuring that the penalty has minimal effect on the optimality of subsequent iterations. Consequently, the proposed method efficiently converges to a near-globally optimal rank-1 solution to the NP-hard MMF and QoS multicast problems.

\begin{figure}
\centering
\setlength{\abovecaptionskip}{0.33cm plus 0pt minus 0pt}
\includegraphics[scale=0.465]{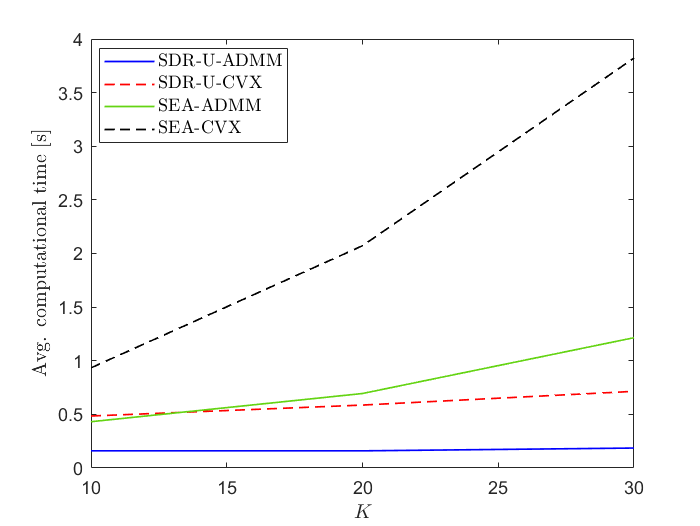}
\caption{Average runtime against $K$ for the MMF problem.}
\label{MMF_avg_runtimes}
\end{figure}

\begin{figure}
\centering
\setlength{\abovecaptionskip}{0.33cm plus 0pt minus 0pt}
\includegraphics[scale=0.465]{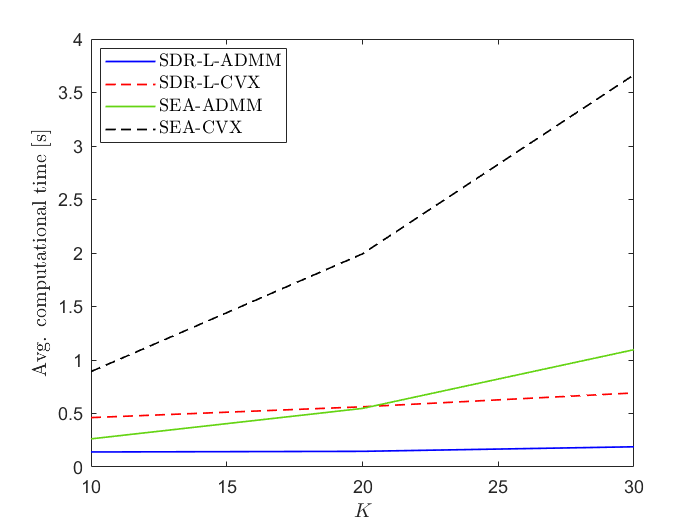}
\caption{Average runtime against $K$ for the QoS problem.}
\label{QoS_avg_runtimes}
\end{figure}

Since order expressions offer limited insight into an algorithm’s complexity for practical problem sizes, owing to the absence of scaling factors, lower-order terms, and their reliance on worst-case bounds that may lead to misleading conclusions, we complement the theoretical complexity analysis with empirical measurements of the average runtimes of each algorithm. Figs. \ref{MMF_avg_runtimes} and \ref{QoS_avg_runtimes} plot the average runtime of the proposed SEA-ADMM, SEA-CVX, and the SDR upper/lower bounds for the MMF and QoS multicast problems, respectively. It can be seen that SEA-ADMM offers tremendous savings in computational requirements when compared to solving the optimization problems with higher-order interior point methods in CVX. Precisely, SEA-ADMM is able to provide between $55\,\%$-$70\,\%$ reduction in runtime over its CVX counterpart for the MMF problem, with the reduction in runtime increasing with $K$. For the QoS problem, SEA-ADMM achieves more than $70\,\%$ decrease in the required computational time for all simulated values of $K$. This demonstrates that the SEA-ADMM algorithm significantly improves the computational requirements for multicast beamforming optimization without any performance loss.  

\subsection{Special Case: Sum-Power Minimization}

In this section, we provide a comparative analysis of the proposed SEA-ADMM algorithm to the prior state-of-the-art ADMM-based SCA algorithm in \cite{zhang2023ultra} developed for the QoS problem, hereafter referred to as ``ASCA". Since the ASCA algorithm considers only a single-cell setup and the extension to a cell-free network with multi-antenna APs and per-AP power budgets is non-trivial, the numerical comparison considers a simplified special case with a single sum-power minimization objective. That is, the proposed SEA-ADMM algorithm can handle the more general cell-free network setting, unlike the ASCA benchmark, in addition to its new approach.

When having a single sum-power minimization objective, the dual variable $\mathbf{z}$ corresponding to the per-AP power constraints becomes redundant. The SEA-ADMM algorithm then simplifies in this case to a similar variant to the one proposed in the conference version of this paper \cite{ADMM_conference_paper}. The SEA-ADMM penalty parameter is set to $\rho = 1$, the inner ADMM penalty parameters corresponding to the $K$ constraints on $\mathbf{y}$ are set to $\mu_k^s = 2 \times 10^6, \forall k$. The rest of the SEA-ADMM parameters are the same as described for the general case.

\begin{figure}
\centering
\setlength{\abovecaptionskip}{0.33cm plus 0pt minus 0pt}
\includegraphics[scale=0.465]{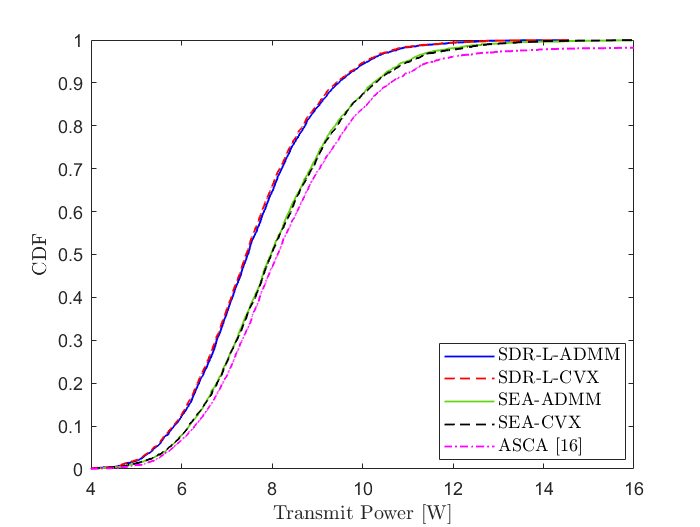}
\caption{CDF of the total transmit power, $K = 30$.}
\label{QoS_CDF2}
\end{figure}

Fig. \ref{QoS_CDF2} plots the CDF of the total transmit power with the sum-power minimization objective for the SEA-ADMM, SEA-CVX and the ADMM- and CVX-based SDR lower bound for $K = 30$ UEs. The prior state-of-the-art ASCA algorithm is shown for comparison. Similar to the more demanding general case described in the previous section, the proposed low-complexity SEA-ADMM algorithm attains the same performance as its CVX-based counterpart. Moreover, a small reduction in the required transmit power is achieved over the ASCA algorithm. It is also worth noting that the ASCA algorithm failed to produce a local optimal solution for about $2\,\%$ of the samples in this setup. The reason lies in the difficulty of finding an effective feasible initial point, which is necessary for SCA-based algorithms, unlike our proposed SEA-ADMM algorithm, which does not require an initial feasible solution. We have observed that this issue becomes even more severe when the problem size grows large, reaching up to a $10\,\%$ failure rate for ADMM-based SCA in some cases.

\begin{figure}
\centering
\setlength{\abovecaptionskip}{0.33cm plus 0pt minus 0pt}
\includegraphics[scale=0.465]{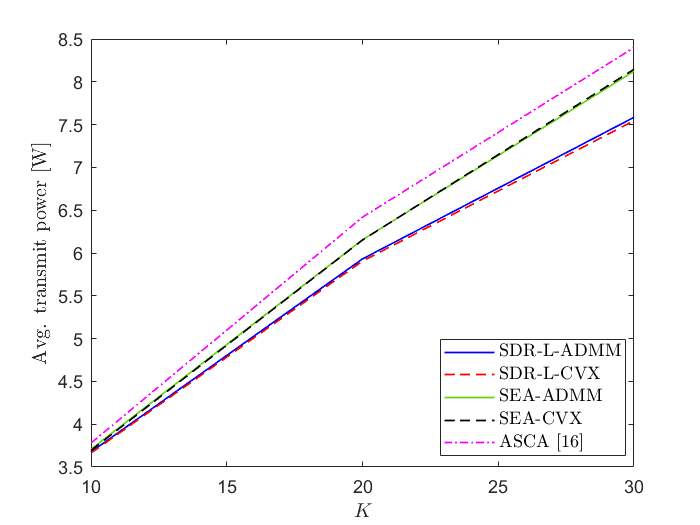}
\caption{Average total transmit power against $K$.}
\label{QoS_avg_power_K}
\end{figure}

In Fig. \ref{QoS_avg_power_K}, we plot the average total transmit power against the number of UEs $K$ for the proposed algorithm and benchmarks. The required transmit power by the network is seen to increase with increasing $K$, a logical outcome of having to satisfy the target SNR for more UEs. The proposed SEA-ADMM algorithm maintains the same performance as SEA-CVX for all values of $K$. Further, we can see an increasing performance improvement for SEA-ADMM compared to the ASCA benchmark as $K$ increases. The reason is that the number of non-convex SNR constraints that are approximated by the SCA method is equal to $K$, degrading its performance as $K$ increases.

\begin{figure}
\centering
\setlength{\abovecaptionskip}{0.33cm plus 0pt minus 0pt}
\includegraphics[scale=0.465]{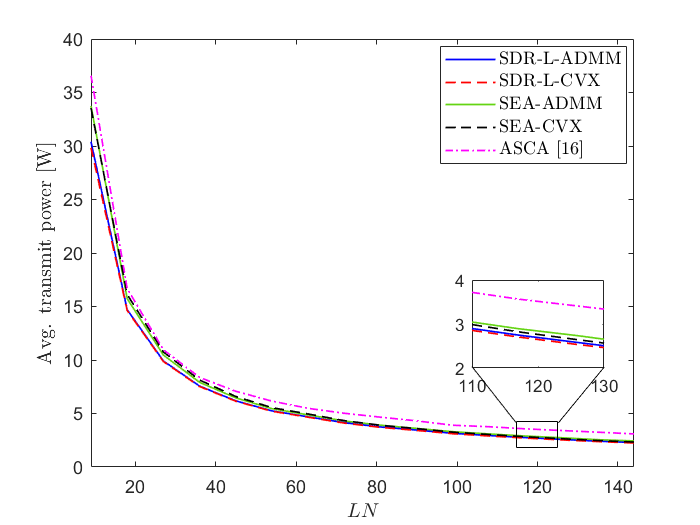}
\caption{Average total transmit power against $LN$, $K = 30$.}
\label{QoS_avg_power_N}
\end{figure}

In Fig. \ref{QoS_avg_power_N}, the average total transmit power is plotted against the total number of antennas $LN$ for the SEA-ADMM algorithm and benchmarks. We highlight that the number of APs is fixed to $L = 9$. It can be seen that as the total number of antennas increases, the required transmit power by the network decreases due to improved beamforming capability. Moreover, the SEA-ADMM is able to keep the same performance as its more complex CVX counterpart for all simulated numbers of antennas. When the total number of antennas is small, around $7\,\%$ increase in transmit power is required by the ASCA algorithm compared to SEA-ADMM. This is associated with the need for more refinement to the multicast beamforming vector when the total number of antennas is much smaller than the number of UEs, emphasizing the performance gap between different approaches. On the other hand, as the problem size grows, the optimality gap of the local optimum solution provided by the ASCA algorithm to the global optimum increases. This is clearly seen as SEA-ADMM is able to achieve approximately the same required transmit power as that of the SDR lower bound, whereas the ASCA algorithm needs up to $25\,\%$ increase in the average transmit power to achieve the same target SNR.

\begin{figure}
\centering
\setlength{\abovecaptionskip}{0.33cm plus 0pt minus 0pt}
\includegraphics[scale=0.465]{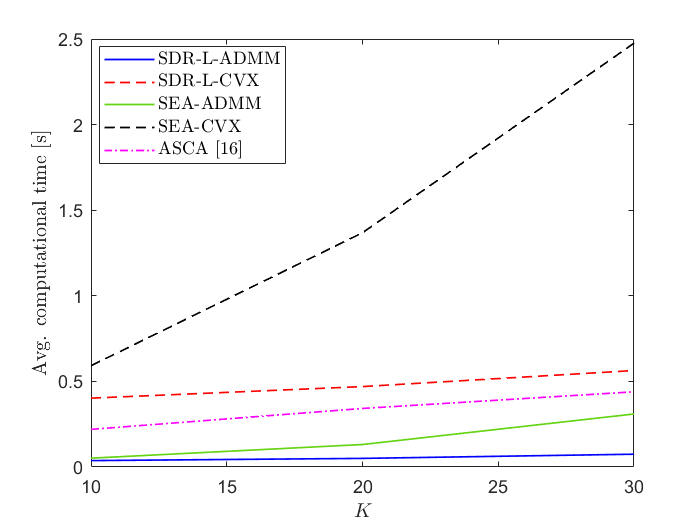}
\caption{Average runtime against $K$.}
\label{QoS_avg_runtimes_K}
\end{figure}

\begin{figure}
\centering
\setlength{\abovecaptionskip}{0.33cm plus 0pt minus 0pt}
\includegraphics[scale=0.465]{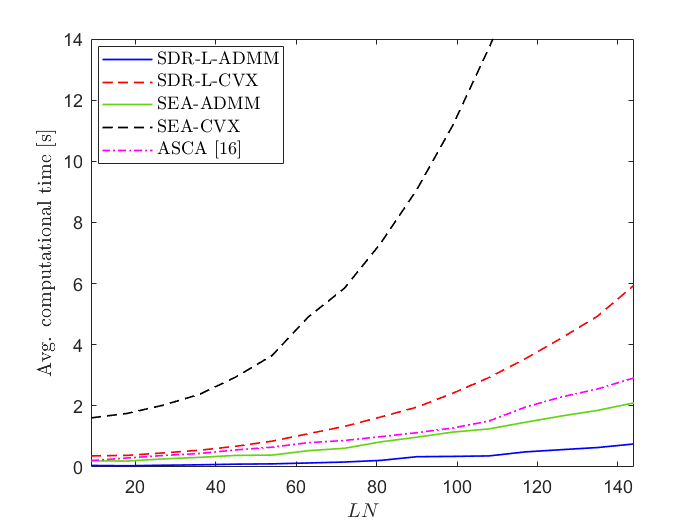}
\caption{Average runtime against $LN$, $K = 30$.}
\label{QoS_avg_runtimes_N}
\end{figure}

Figs. \ref{QoS_avg_runtimes_K} and \ref{QoS_avg_runtimes_N} plot the average runtime for the proposed SEA-ADMM, SEA-CVX, and corresponding lower bounds against $K$ and $LN$, respectively. For the special case of sum-power minimization, it can be seen that roughly $10$ times reduction in computational time is achievable by the proposed SEA-ADMM algorithm compared to its CVX counterpart. In comparison to the ASCA benchmark, the SEA-ADMM algorithm maintains a lower runtime for all simulated scenarios. The reduction in runtime for SEA-ADMM is particularly large when the number of UEs is relatively small, reaching up to $60\,\%$ lower runtime compared to the ASCA algorithm. When comparing the two algorithms for different $LN$ and $K = 30$ UEs, SEA-ADMM is able to achieve up to $33\,\%$ reduction in runtime compared to the ASCA algorithm. These results demonstrate the superiority and scalability of the proposed approach in different network settings as compared to the state-of-the-art, besides its wider domain of applicability.

\subsection{Effect of CSI Uncertainty}

To evaluate the effect of CSI uncertainty, in the following, we assume AP $l$ has an estimate of the channel to UE $k$ as
\begin{equation}
    \hat{\mathbf{h}}_{kl} = \sqrt{1 - \tau_e^2}\mathbf{h}_{kl} + \tau_e\mathbf{e}_{kl},
    \label{CSI_error_model}
\end{equation}
where $\mathbf{e}_{kl} \sim \mathcal{N}_{\mathbb{C}}(\mathbf{0}, \mathbf{R}_{kl})$ is the independent CSI error and $0 \leq \tau_e \leq 1$ determines the quality of the available CSI.

\begin{figure}
\centering
\setlength{\abovecaptionskip}{0.33cm plus 0pt minus 0pt}
\includegraphics[scale=0.465]{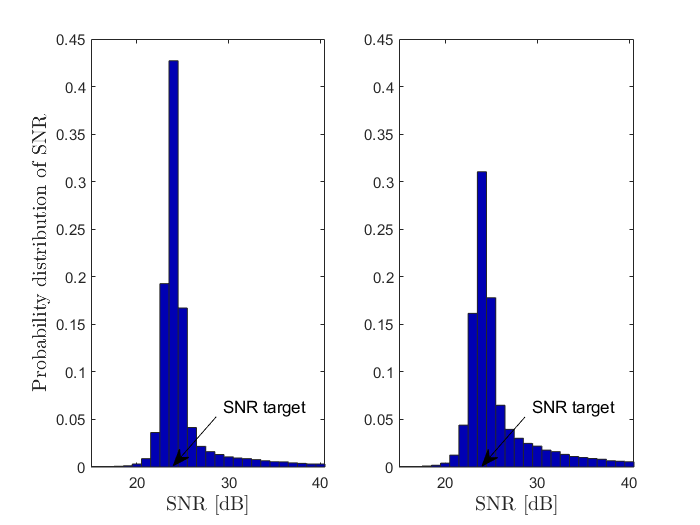}
\caption{Distribution of the achievable SNR with imperfect CSI for $K = 10$ (left) and $K = 30$ (right).}
\label{CSI_error_fig}
\end{figure}

Fig. \ref{CSI_error_fig} plots the probability distribution of the achievable SNR under the CSI error model in \eqref{CSI_error_model} with $\tau_e = 0.2$ for $K = 10$ and $K = 30$ UEs. A direct approach to combat the effect of CSI uncertainty is to consider a CSI error margin. The figure shows that for a $3~\mathrm{dB}$ error margin, the outage probability in which a UEs' SNR falls below the error margin is about $1\,\%$ and $1.6\,\%$ for $K = 10$ and $K = 30$ UEs, respectively. This error margin can be judiciously adjusted to achieve a certain target outage probability. We highlight that the effect of CSI error is independent of the choice of the optimization algorithm.

\section{Conclusions And Future Work} \label{conc}

In this work, the MMF and QoS multicast beamforming optimization problems have been revisited within the framework of cell-free massive MIMO. We introduced a novel iterative optimization approach capable of achieving a near-globally optimal rank-1 beamforming solution to this NP-hard problem. The proposed method builds upon the SEA to extract a rank-1 solution from the SDP output and incorporates a novel problem reformulation and penalization strategy specifically designed for the SEA-ADMM framework. The proposed SEA-ADMM algorithm is evaluated against state-of-the-art SDR- and SCA-based multicast beamforming optimization procedures. Numerical results demonstrate substantial gains in both performance and computational efficiency while utilizing the same platform and software to solve optimization problems. These improvements stem from the SEA-ADMM algorithm’s ability to eliminate higher-rank solutions while imposing minimal impact on the optimal achievable SE, whereas previous SCA methods rely on approximations that degrade the performance, especially as the number of UEs increases. Moreover, the results highlight the robustness of the SEA-ADMM algorithm where it maintains its performance across different simulation scenarios. The proposed framework is broadly applicable to optimization problems where SDR is employed and low-rank solutions are desired, which is a wide class of optimization problems. Compared to standard SDP solvers, SEA-ADMM attains similar performance with approximately $70\,\%$ reduction in computational time, bringing real-time optimal multicast beamforming a step closer.

Future extensions may include considering multi-group multicasting, where the inter-group interference makes the multicast beamforming optimization problem more challenging, particularly for the MMF objective. For the multi-group multicast QoS problem, an extension to the SEA-ADMM algorithm is possible. The common approach to solve the MMF problem is then to rely on solving a sequence of the relaxed QoS problem, as detailed in \cite{karipidis2008quality,hsu2017joint,zaher2026cell}. Moreover, combining the proposed new solution approach with robust multicast beamforming design can mitigate the effect of channel uncertainties. In general, such designs tackle the multicast beamforming problem considering a bounded CSI error model by utilizing worst-case SINR reformulations and the S-lemma, as can be found in \cite{hsu2017joint,mohamadi2022low}, and references therein. Finally, one can train deep learning models to attempt to achieve similar performance for lower complexity.

\section*{Appendix A\\Proof of Theorem 1}
We start by formulating the Lagrangian of problem \eqref{MMF_problem3} as
\begin{equation}
\begin{split}
    L_{\textrm{MMF}}\left(\mathbf{W}, t, \mathbf{y}, \mathbf{z}, \mathbf{S}\right) = &-t - \langle \mathbf{y}, \mathcal{H}\left(\mathbf{W}\right) - \mathbf{1}t\rangle  \\
    &+ \langle\mathbf{z}, \mathcal{D}\left(\mathbf{W}\right)  - \mathbf{p}\rangle - \langle\mathbf{S}, \mathbf{W}\rangle,
\end{split}
\label{MMF_lag}
\end{equation}
Taking the derivative of the Lagrangian with respect to $\mathbf{W}$ and equating it to $\mathbf{0}$, we get
\begin{equation}
    \mathcal{H}^H\left(\mathbf{y}\right) + \mathbf{S} = \mathcal{D}^H\left(\mathbf{z}\right).
    \label{MMFcondition}
\end{equation}
Further, equating the derivative of the Lagrangian with respect to $t$ to $0$, we have
\begin{equation}
    -1 + \mathbf{y}^T\mathbf{1} = 0 \quad\Rightarrow\quad \norm{\mathbf{y}}_1 = 1. \label{MMFcondition2}
\end{equation}
Plugging the conditions in \eqref{MMFcondition} and \eqref{MMFcondition2} back in \eqref{MMF_lag}, the dual function becomes $g\left(\mathbf{W}, t, \mathbf{y}, \mathbf{z}, \mathbf{S}\right) = -\mathbf{z}^T\mathbf{p}$ when the conditions are satisfied. The dual SDP problem to the MMF problem is then the maximum of the dual function subject to the conditions \eqref{MMFcondition} and \eqref{MMFcondition2}, which can be reformulated as \eqref{dualMMF} thereby completing the proof.

\section*{Appendix B\\Proof of Theorem 2}

We start by vectorizing the matrix terms in the objective of problem \eqref{yz_update_problem} using the fact that for a matrix $\mathbf{A}$, $\norm{\mathbf{A}}_F^2 = \norm{\textrm{vec}(\mathbf{A})}_2^2$. Accordingly, the objective in problem \eqref{yz_update_problem} can be reformulated as
\begin{equation}
\begin{split}
    \mathbf{z}^T\mathbf{p} &+ \frac{\rho}{2}\norm{\mathcal{H}^H\left(\mathbf{y}\right) + \mathbf{S}^i - \mathcal{D}^H\left(\mathbf{z}\right) + \overbar{\mathbf{W}}^i}_F^2\\
    &= \mathbf{z}^T\mathbf{p} + \frac{\rho}{2}\norm{\mathbf{H}^H\mathbf{y} - \mathbf{D}^H\mathbf{z} + \mathbf{r}}_2^2\\
    &= \mathbf{z}^T\mathbf{p} + \frac{\rho}{2}\left(\mathbf{H}^H\mathbf{y} - \mathbf{D}^H\mathbf{z} + \mathbf{r}\right)^H\left(\mathbf{H}^H\mathbf{y} - \mathbf{D}^H\mathbf{z} + \mathbf{r}\right)\\
    &= \mathbf{z}^T\mathbf{p} + \frac{\rho}{2}\Bigl(\mathbf{y}^T\mathbf{H}\mathbf{H}^H\mathbf{y} + \mathbf{z}^T\mathbf{D}\mathbf{D}^H\mathbf{z} - 2\Re\{\mathbf{z}^T\mathbf{D}\mathbf{H}^H\mathbf{y}\}\\ 
    &\hspace{10pt}+ 2\Re\{\mathbf{y}^T\mathbf{H}\mathbf{r}\} - 2\Re\{\mathbf{z}^T\mathbf{D}\mathbf{r}\} + \mathbf{r}^H\mathbf{r}\Bigr)\\
    &= \frac{\rho}{2}\begin{bmatrix}
    \mathbf{y}\\\mathbf{z}
    \end{bmatrix}^T\begin{bmatrix}
    \mathbf{HH}^H & -\mathbf{HD}^H\\-\mathbf{DH}^H & \mathbf{DD}^H
    \end{bmatrix}\begin{bmatrix}
    \mathbf{y}\\\mathbf{z}
    \end{bmatrix}\\
    &\hspace{10pt}+ \begin{bmatrix}
    \mathbf{y}\\\mathbf{z}
    \end{bmatrix}^T\begin{bmatrix}
    \rho\hspace{1pt}\Re\{\mathbf{Hr}\}\\
    \mathbf{p}-\rho\hspace{1pt}\Re\{\mathbf{Dr}\}
    \end{bmatrix} + \frac{\rho}{2}\mathbf{r}^H\mathbf{r}\\
    &= \frac{1}{2}\mathbf{x}^T\mathbf{Qx} + \mathbf{c}^T\mathbf{x} + \frac{\rho}{2}\mathbf{r}^H\mathbf{r}.
\end{split}
\end{equation}
Ignoring the last term, which is independent of the optimization variables $\{\mathbf{y}, \mathbf{z}\}$, problem \eqref{yz_update_problem} can be rewritten in standard QP form as given in problem \eqref{QP}.

\section*{Appendix C\\Sorting-Based Simplex Projection}

\begin{algorithm}
\caption{Simplex Projection of a Vector}
\noindent\textbf{Input:} $\mathbf{v} \in \mathbb{R}^n$, $\tau > 0$.
\begin{algorithmic}[1]
\STATE Set $\mathbf{u} \leftarrow$ Sorted $\mathbf{v}$ in descending order.
\STATE Compute $\mathbf{c} \leftarrow$ Cumulative sum of the sorted vector $\mathbf{u}$.
\STATE Find $\eta = \max \Big\{ j \in \{1,\dots,n\} \;|\; u_j - \frac{c_j - \tau}{j} > 0 \Big\}$.
\STATE Compute threshold: $\theta = \frac{c_\eta - \tau}{\eta}$.
\STATE Compute projection: $q_i = \max(v_i - \theta, 0), \quad i = 1,\dots,n$.
\end{algorithmic}
\noindent\textbf{Output:} Projected vector $\mathbf{q} \in \mathbb{R}^n$ on the simplex.
\end{algorithm}

\section*{References}
\renewcommand{\refname}{ \vspace{-\baselineskip}\vspace{-1.1mm} }
\bibliographystyle{ieeetr}
\bibliography{papercites}

\begin{thebibliography}{10}

\bibitem{hsu2017joint}
G.-W. Hsu, B.~Liu, H.-H. Wang, and H.-J. Su, ``Joint beamforming for multicell multigroup multicast with per-cell power constraints,'' {\em IEEE Transactions on Vehicular Technology}, vol.~66, no.~5, pp.~4044--4058, 2017.

\bibitem{zaher2023soft}
M.~Zaher, E.~Bj{\"o}rnson, and M.~Petrova, ``Soft handover procedures in {mmWave} cell-free massive {MIMO} networks,'' {\em IEEE Transactions on Wireless Communications}, vol.~23, no.~6, pp.~6124--6138, 2024.

\bibitem{bjornson2019making}
E.~Bj{\"o}rnson and L.~Sanguinetti, ``Making cell-free massive {MIMO} competitive with {MMSE} processing and centralized implementation,'' {\em IEEE Transactions on Wireless Communications}, vol.~19, no.~1, pp.~77--90, 2019.

\bibitem{demir2021foundations}
{\"O}.~T. Demir, E.~Bj{\"o}rnson, and L.~Sanguinetti, ``Foundations of user-centric cell-free massive {MIMO},'' {\em Foundations and Trends{\textregistered} in Signal Processing}, vol.~14, no.~3-4, pp.~162--472, 2021.

\bibitem{zaher2023learning}
M.~Zaher, {\"O}.~T. Demir, E.~Bj{\"o}rnson, and M.~Petrova, ``Learning-based downlink power allocation in cell-free massive {MIMO} systems,'' {\em IEEE Transactions on Wireless Communications}, vol.~22, no.~1, pp.~174--188, 2023.

\bibitem{karipidis2008quality}
E.~Karipidis, N.~D. Sidiropoulos, and Z.-Q. Luo, ``Quality of service and max-min fair transmit beamforming to multiple cochannel multicast groups,'' {\em IEEE Transactions on Signal Processing}, vol.~56, no.~3, pp.~1268--1279, 2008.

\bibitem{sidiropoulos2006transmit}
N.~D. Sidiropoulos, T.~N. Davidson, and Z.-Q. Luo, ``Transmit beamforming for physical-layer multicasting,'' {\em IEEE Transactions on Signal Processing}, vol.~54, no.~6, pp.~2239--2251, 2006.

\bibitem{ashraphijuo2017multicast}
M.~Ashraphijuo, X.~Wang, and M.~Tao, ``Multicast beamforming design in multicell networks with successive group decoding,'' {\em IEEE Transactions on Wireless Communications}, vol.~16, no.~6, pp.~3492--3506, 2017.

\bibitem{zhou2017coordinated}
L.~Zhou, L.~Zheng, X.~Wang, W.~Jiang, and W.~Luo, ``Coordinated multicell multicast beamforming based on manifold optimization,'' {\em IEEE Communications Letters}, vol.~21, no.~7, pp.~1673--1676, 2017.

\bibitem{xiang2014massive}
Z.~Xiang, M.~Tao, and X.~Wang, ``Massive {MIMO} multicasting in noncooperative cellular networks,'' {\em IEEE Journal on Selected Areas in Communications}, vol.~32, no.~6, pp.~1180--1193, 2014.

\bibitem{sadeghi2017reducing}
M.~Sadeghi, L.~Sanguinetti, R.~Couillet, and C.~Yuen, ``Reducing the computational complexity of multicasting in large-scale antenna systems,'' {\em IEEE Transactions on Wireless Communications}, vol.~16, no.~5, pp.~2963--2975, 2017.

\bibitem{sadeghi2017max}
M.~Sadeghi, E.~Bj{\"o}rnson, E.~G. Larsson, C.~Yuen, and T.~L. Marzetta, ``Max--min fair transmit precoding for multi-group multicasting in massive {MIMO},'' {\em IEEE Transactions on Wireless Communications}, vol.~17, no.~2, pp.~1358--1373, 2017.

\bibitem{de2022user}
A.~de~la Fuente, G.~Interdonato, and G.~Araniti, ``User subgrouping and power control for multicast massive {MIMO} over spatially correlated channels,'' {\em IEEE Transactions on Broadcasting}, vol.~68, no.~4, pp.~834--847, 2022.

\bibitem{li2022spectral}
J.~Li, Q.~Pan, Z.~Wu, P.~Zhu, D.~Wang, and X.~You, ``Spectral efficiency of unicast and multigroup multicast transmission in cell-free distributed massive {MIMO} systems,'' {\em IEEE Transactions on Vehicular Technology}, vol.~71, no.~12, pp.~12826--12839, 2022.

\bibitem{dong2020multi}
M.~Dong and Q.~Wang, ``Multi-group multicast beamforming: Optimal structure and efficient algorithms,'' {\em IEEE Transactions on Signal Processing}, vol.~68, pp.~3738--3753, 2020.

\bibitem{zhang2023ultra}
C.~Zhang, M.~Dong, and B.~Liang, ``Ultra-low-complexity algorithms with structurally optimal multi-group multicast beamforming in large-scale systems,'' {\em IEEE Transactions on Signal Processing}, vol.~71, pp.~1626--1641, 2023.

\bibitem{dartmann2011low}
G.~Dartmann, X.~Gong, and G.~Ascheid, ``Low complexity cooperative multicast beamforming in multiuser multicell downlink networks,'' in {\em Proc. 6th International ICST Conference on Cognitive Radio Oriented Wireless Networks and Communications (CROWNCOM)}, pp.~370--374, 2011.

\bibitem{tran2013conic}
L.-N. Tran, M.~F. Hanif, and M.~Juntti, ``A conic quadratic programming approach to physical layer multicasting for large-scale antenna arrays,'' {\em IEEE Signal Processing Letters}, vol.~21, no.~1, pp.~114--117, 2013.

\bibitem{globecom_paper}
M.~Zaher, E.~Bj{\"o}rnson, and M.~Petrova, ``Near-optimal cell-free beamforming for physical layer multigroup multicasting,'' in {\em Proc. IEEE Global Communications Conference (GLOBECOM)}, pp.~1--6, 2024.

\bibitem{zaher2026cell}
M.~Zaher, E.~Bj{\"o}rnson, and M.~Petrova, ``Cell-free beamforming design for physical layer multigroup multicasting,'' {\em IEEE Transactions on Wireless Communications}, vol.~25, pp.~5262--5274, 2026.

\bibitem{wang2016weighted}
J.~Wang, D.~Wang, and Y.~Liu, ``Weighted sum rate-based coordinated beamforming in multi-cell multicast networks,'' {\em IEEE Communications Letters}, vol.~20, no.~8, pp.~1567--1570, 2016.

\bibitem{park2008capacity}
S.~Y. Park and D.~J. Love, ``Capacity limits of multiple antenna multicasting using antenna subset selection,'' {\em IEEE Transactions on Signal Processing}, vol.~56, no.~6, pp.~2524--2534, 2008.

\bibitem{chen2017admm}
E.~Chen and M.~Tao, ``{ADMM}-based fast algorithm for multi-group multicast beamforming in large-scale wireless systems,'' {\em IEEE Transactions on Communications}, vol.~65, no.~6, pp.~2685--2698, 2017.

\bibitem{konar2017fast}
A.~Konar and N.~D. Sidiropoulos, ``Fast approximation algorithms for a class of non-convex {QCQP} problems using first-order methods,'' {\em IEEE Transactions on Signal Processing}, vol.~65, no.~13, pp.~3494--3509, 2017.

\bibitem{ibrahim2020fast}
M.~S. Ibrahim, A.~Konar, and N.~D. Sidiropoulos, ``Fast algorithms for joint multicast beamforming and antenna selection in massive {MIMO},'' {\em IEEE Transactions on Signal Processing}, vol.~68, pp.~1897--1909, 2020.

\bibitem{mohamadi2022low}
N.~Mohamadi, M.~Dong, and S.~ShahbazPanahi, ``Low-complexity {ADMM}-based algorithm for robust multi-group multicast beamforming in large-scale systems,'' {\em IEEE Transactions on Signal Processing}, vol.~70, pp.~2046--2061, 2022.

\bibitem{mohamadi2024low}
N.~Mohamadi, M.~Dong, and S.~ShahbazPanahi, ``Low-complexity joint antenna selection and robust multi-group multicast beamforming for massive {MIMO},'' {\em IEEE Transactions on Signal Processing}, vol.~72, pp.~792--808, 2024.

\bibitem{huang2016consensus}
K.~Huang and N.~D. Sidiropoulos, ``Consensus-{ADMM} for general quadratically constrained quadratic programming,'' {\em IEEE Transactions on Signal Processing}, vol.~64, no.~20, pp.~5297--5310, 2016.

\bibitem{ADMM_conference_paper}
M.~Zaher and E.~Bj{\"o}rnson, ``Low-complexity {SDP-ADMM} for physical-layer multicasting in massive {MIMO} systems,'' in {\em Proc. 23rd International Symposium on Modeling and Optimization in Mobile, Ad Hoc, and Wireless Networks (WiOpt)}, pp.~1--7, 2025.

\bibitem{luo2010semidefinite}
Z.-Q. Luo, W.-K. Ma, A.~M.-C. So, Y.~Ye, and S.~Zhang, ``Semidefinite relaxation of quadratic optimization problems,'' {\em IEEE Signal Processing Magazine}, vol.~27, no.~3, pp.~20--34, 2010.

\bibitem{boyd2011distributed}
S.~Boyd, N.~Parikh, E.~Chu, B.~Peleato, J.~Eckstein, {\em et~al.}, ``Distributed optimization and statistical learning via the alternating direction method of multipliers,'' {\em Foundations and Trends{\textregistered} in Machine learning}, vol.~3, no.~1, pp.~1--122, 2011.

\bibitem{duchi2008efficient}
J.~Duchi, S.~Shalev-Shwartz, Y.~Singer, and T.~Chandra, ``Efficient projections onto the l1-ball for learning in high dimensions,'' in {\em Proc. 25th International Conference on Machine Learning}, pp.~272--279, 2008.

\bibitem{bjornson2017book}
E.~Bj{\"o}rnson, J.~Hoydis, and L.~Sanguinetti, ``Massive {MIMO} networks: Spectral, energy, and hardware efficiency,'' {\em Foundations and Trends in Signal Processing}, vol.~11, no.~3-4, pp.~154--655, 2017.

\bibitem{cvx}
M.~Grant and S.~Boyd, ``{CVX}: Matlab software for disciplined convex programming, version 2.1.'' \url{https://cvxr.com/cvx}, Mar. 2014.

\end{thebibliography}

\end{document}